\documentclass[prl,twocolumn,superscriptaddress]{revtex4}
\usepackage{amssymb}
\usepackage{graphicx,amsmath}
\usepackage{bm}
\usepackage{times}

\begin{document}

\title{Multiphoton transitions in Josephson-junction qubits \\
(Review Article)}
\author{S. N. Shevchenko}
\email{sshevchenko@ilt.kharkov.ua} \affiliation{B.Verkin Institute
for Low Temperature Physics and Engineering, Kharkov, Ukraine}
\author{A. N. Omelyanchouk}
\affiliation{B.Verkin Institute for Low Temperature Physics and
Engineering, Kharkov, Ukraine}
\author{E.~Il'ichev}
\affiliation{Institute of Photonic Technology, Jena, Germany}

\begin{abstract}
Two basic physical models, a two-level system and a harmonic oscillator, are
realized on the mesoscopic scale as coupled qubit and resonator. The
realistic system includes moreover the electronics for controlling the
distance between the qubit energy levels and their populations and to read
out the resonator's state, as well as the unavoidable dissipative
environment. Such rich system is interesting both for the study of
fundamental quantum phenomena on the mesoscopic scale and as a promising
system for future electronic devices.

We present recent results for the driven superconducting qubit - resonator
system, where the resonator can be realized as an $LC$ circuit or a
nanomechanical resonator. Most of the results can be described by the
semiclassical theory, where a qubit is treated as a quantum two-level system
coupled to the classical driving field and the classical resonator.
Application of this theory allows to describe many phenomena for the single
and two coupled superconducting qubits, among which are the following: the
equilibrium-state and weak-driving spectroscopy, Sisyphus damping and
amplification, Landau-Zener-St\"{u}ckelberg interferometry, the multiphoton
transitions of both direct and ladder-type character, and creation of the
inverse population for lasing.
\end{abstract}

\date{\today}
\maketitle
\tableofcontents


\section{1. Introduction}

A quantum system, subjected to external driving, can experience resonant
transitions between its energy levels. Conservation of total energy assumes
absorption or emission of several photons of the driving field. Such
multiphoton processes play an important role in atomic and molecular systems
interacting with electromagnetic field \cite{Coh-Tan}. For example, the
multiphoton resonant spectroscopy is one of the methods to probe the
structure of atoms and molecules \cite{Delone}. This technique has the
advantage of observing highly excited states by using relatively low
frequencies. The concept of another application, the multiphoton excitation
microscopy, is based on the multiphoton excitation of the fluorescent dyes
molecules \cite{Xu96, Konig00, Diaspro06}. This technique allows imaging
biochemical objects with high spatial resolution.

Recent development of fabrication and measurement techniques enabled a study
of the wide spectrum of quantum phenomena in superconducting structures.
During the past years it has been clearly shown that specially designed
macroscopic superconducting circuits, which include Josephson junctions,
behave quantum mechanically similar to a quantum particle in a potential
well. Under certain conditions, these objects demonstrate the coherent
superposition between their macroscopically distinct quantum states. It is
important to note that this is a pure quantum effect which has no classical
analogue and can be used for a number of intrigued applications. If the
circuit's dynamics can be described in the frame of the two-level
approximation, such two-level quantum system is called a qubit. The advance
in the study of different phenomena in superconducting qubits can be found
in the reviews \cite{Makhlin01, Devoret04, Wendin07, Clarke08, You11}.

In general, superconducting Josephson circuits can be described as
multilevel quantum systems. By analogy, such systems are called artificial
atoms, while coupled qubits systems behave as artificial molecules. An
interesting problem is how phenomena, known from atomic physics, will appear
for these artificial atoms and molecules. Note that the following features
differ these mesoscopic-size quantum systems from their microscopic
counterparts: a high level of controllability by electronic means, coupling
to the macroscopic-size read-out devices, and unavoidable dissipative
environment.

For characterization and controlling the states of superconducting qubits
the one-photon spectroscopy was done by using relatively weak driving \cite%
{Nakamura99, Friedman00, vanderWal00, Vion02, Yu02, Martinis02, Chiorescu03,
Born04}. Matching of the ground and higher states with the one-photon energy
was exploited to probe the upper levels of the Josephson-junction circuits
\cite{Claudon04, Berns08, Neely09, Nori09, Jirari09, Sillanpaa09, Joo10,
Du10a}. With increasing driving power, the multiphoton excitations were used
to study the features of the artificial atoms both for the two-level
dynamics \cite{Nakamura01, Wallraff03, Saito04, Shnyrkov06, Shnyrkov09}\ and
when the upper levels were involved \cite{Yu05, Strauch07, Dutta08, Wang10,
Bushev10}. For strong driving, the width of the resonance lines periodically
tends to zero, which can be described as the destructive Landau-Zener-St\"{u}%
ckelberg interference \cite{Shevchenko10}. Respective interferograms
displayed double-periodical dependence of the upper-level occupation
probability on the energy bias and the driving amplitude \cite{Oliver05,
Sillanpaa06, Wilson07, Izmalkov08, Sun09, LaHaye09}.

Two and more coupled qubits can be treated as artificial molecules. Being
excited by a resonant microwave field, they display one-photon transitions
\cite{Pashkin03, Berkley03, Majer05, Steffen06, Plantenberg07, Fay08,
DiCarlo09, Altomare10}. Alternatively, at smaller frequencies, the two-qubit
systems can experience multiphoton transitions \cite{Leek09, Ilichev10,
Temchenko11, Satanin12}.

In this article we review the observations of the multiphoton transitions in
single and coupled superconducting qubits probed by a classical resonator,
and also we present the respective theory. Having the purpose of presenting
and describing specific results for the multiphoton transitions, our
consideration is limited to the Josephson-junction qubits. We note however
that similar phenomena can be studied in different quantum objects, which
can be described as two- or multi-level systems, such as quantum wires and
dots \cite{Ho09, Ribeiro10, Brataas11, Dovzhenko11, Gaudreau12}, nitrogen
vacancy centers in diamond \cite{Childress10, Huang11}, ultracold atoms \cite%
{Zenesini10, Plotz10, Zhang11}, nanomechanical and optomechanical setups
\cite{Heinrich10, Chotorlishvili11, Miladinovic11}, electronic spin systems,
two-dimensional electron gas, and graphene \cite{Bertaina11, Hatke11,
Avetissian11}.

The paper is organized as follows. In Sec.~2 we use the method of an
asymptotic expansion for the qubit-resonator system in order to obtain the
resonator characteristics. This formalism allows us to separate the dynamics
of the relatively slow resonator and fast qubit. Then, in Sec.~3, we
consider the multiphoton dynamics of an isolated two-level system. Later the
formulas of those two sections will be applied for the description of the
experimentally observed multiphoton excitations in single qubits (Sec.~4)
and in coupled qubits systems (Sec.~5).

\section{2. Semiclassical theory of the qubit-resonator system}

For characterization of a quantum system different techniques can be
applied. One of the possible solutions is to use the so-called parametric
transducer~\cite{Braginsky92}. A key element in any parametric transducer is
an optical or a radio-frequency auto-oscillator. A transducer, coupled to
the quantum system of interest, is constructed so, that quantum system
dynamics causes a change of the phase or/and the amplitude of its
oscillations. A phase (amplitude) shift provides information about the
dynamics of a quantum system. In particular, for probing the qubit's state,
several types of oscillators have been already used: an $LC$ tank circuit~%
\cite{Ilichev02, Ilichev09}, a nanomechanical one~\cite{Irish03, LaHaye09},
and a transmission line resonator \cite{Blais04, Schuster05}. If the
resonator quantization energy $\hbar \omega _{\mathrm{p}}$ is smaller than
the thermal excitation energy $k_{\mathrm{B}}T$, the resonator can be
considered as a classical oscillator. Then the qubit-resonator system can be
treated semiclassically: here a qubits quantum system is driven by a
classical field and probed by a classical oscillator. It is important to
note that the similar approach is well known in quantum optics - many
phenomena in the atom-light system can be described by making use of this
semiclassical model~\cite{Delone}.

In this work we present the semiclassical description of some
observed effects for the resonator-qubits systems. We will not
consider here the situation of coupling the qubits systems to a
high-frequency resonator, which can be realized as a transmission
line resonator. The quantum properties of this qubit-resonator
system are not described by the semiclassical model. For recent
works in this field see e.g. \cite{Abdumalikov08, Oelsner09,
Omelyanchouk10, Ashhab10, Niemczyk10} and references therein and
also Refs. \cite{Bishop08, Niemczyk09, Fink09}, where the
multiphoton excitations were used to drive transitions between the
multiple energy levels of the qubit-resonator system in the strong
coupling regime.

Another note here should be made about the term \textquotedblleft
multiphoton processes\textquotedblright . In the context of the
semiclassical approach, it relates to the energy of several photons which is
absorbed or emitted by the quantum system. In the broader sense the term
\textquotedblleft multiphoton\textquotedblright\ can relate to other
processes employing the quantum nature of the electromagnetic field, see
Ref. \cite{Pan11} for a review of the non-classical phenomena in entangled
multi-photon systems.

This section is devoted to the properties of the qubit-resonator system. It
will be shown that in the frame of the semiclassical approach the influence
of the qubit on the resonator can be described by the \textquotedblleft
renormalization\textquotedblright\ of the oscillator constants. For instance
for a mechanical resonator it can be quantified by introducing the
equivalent qubit's-state-dependent elasticity coefficient and damping
factor. In the case of inductive/capacitive coupling, the qubit's impact on
the resonator can be described by introducing the qubit's-state-dependent
effective inductance/capacitance, while the losses can be described by the
effective resistance. For concreteness, we will consider two realistic
systems: the flux qubit inductively coupled to the tank circuit~\cite%
{ShevchenkoPG08} and the charge qubit capacitively coupled to the
nanomechanical resonator~\cite{Shevchenko11}.

\subsection{Krylov-Bogolyubov formalism for qubit-resonator system}

\begin {widetext}

\begin{figure}[t]
\includegraphics[width=10 cm]{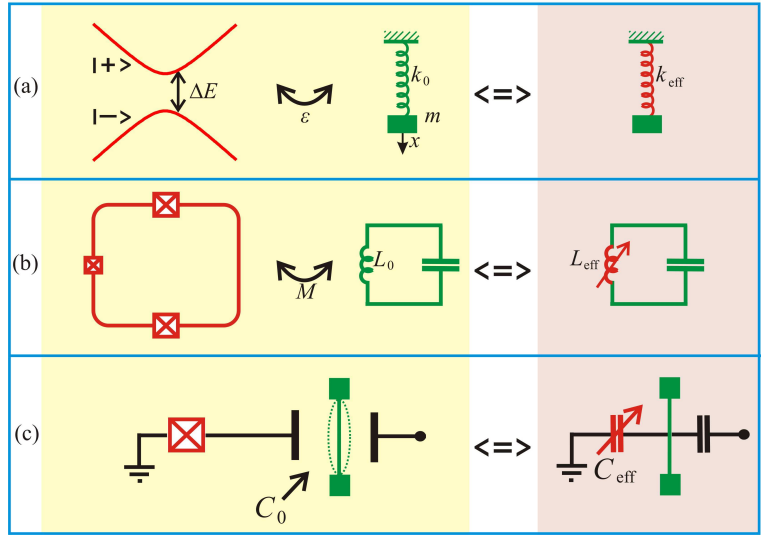}
\caption{(Color online) \textbf{Qubit (quantum two-level system)
coupled to a classical resonator.} (a) Schematic diagram of the
model qubit-resonator system. The qubit is represented by the two-level system
with the two states, $\left\vert -\right\rangle $ and $\left\vert
+\right\rangle \,$, and with the energy difference $\Delta E$. The
resonator is demonstrated as the spring oscillator with the
elasticity coefficient $k_{0}$. As described in the main text,
influence of the qubit on the resonator can be described by
introducing the effective elasticity coefficient
$k_{\mathrm{eff}}$, which includes the qubit's-state-dependent
(or, parametric, for brevity) elasticity coefficient
$k_{\mathrm{q}}$. (b) The flux qubit coupled via the mutual
inductance $M$ to the $LC$ resonator. This can be described by
introducing effective qubit's-state-dependent inductance
$L_{\mathrm{eff}}$, which includes the parametric inductance
$L_{\mathrm{q}}$ in parallel to the tank's inductance $L_{0}$. (c)
The impact of the charge qubit on the nanomechanical resonator's
state can be described by introducing the effective
qubit's-state-dependent capacitance $C_{\mathrm{eff}}$, which
includes the parametric capacitance $C_{\mathrm{q}}$ in parallel to
the resonator's capacitance $C_{0}$.} \label{Fig:effectiveL&C}
\end{figure}

\end {widetext}

First let us consider the mechanical resonator as a spring with the
elasticity $k_{0}$, the damping factor $\lambda _{0}$ (which is assumed to
be small), and loaded with mass $m$, as shown in Fig.~\ref{Fig:effectiveL&C}%
(a). The oscillator has eigenfrequency $\omega _{0}=\sqrt{k_{0}/m}$ and the
quality factor $Q_{0}=m\omega _{0}/\lambda _{0}$. Its state is influenced by
the qubit through the force $\epsilon F_{\mathrm{q}}$ and is driven by the
probe periodical force $\epsilon F_{\mathrm{p}}\sin \omega _{\mathrm{p}}t$.
Here the small parameter $\epsilon $ is introduced explicitly to emphasize
the small qubit-resonator coupling as well as the amplitude of the external
harmonic force $\epsilon F_{\mathrm{p}}$, which enables us to make use of
the asymptotic expansion method. The external nonlinear force is assumed to
depend on the variable $x$ and its derivative only, $F_{\mathrm{q}}=F_{%
\mathrm{q}}(x,dx/dt)$.

The displacement $x$ is the solution of the motion equation
\begin{equation}
m\frac{d^{2}x}{dt^{2}}+\lambda _{0}\frac{dx}{dt}+k_{0}x=\epsilon F_{\mathrm{q%
}}\left( x,\frac{dx}{dt}\right) +\epsilon F_{\mathrm{p}}\sin \omega _{%
\mathrm{p}}t.  \label{eq4x}
\end{equation}%
The oscillations in the nonlinear system described by Eq.~(\ref{eq4x}) can
be reduced to oscillations in an \textit{equivalent linear system} by making
use of the Krylov-Bogolyubov technique of asymptotic expansion~\cite%
{Bogolyubov}. Specifically, in the first-order approximation with respect to
a qubit-resonator coupling parameter and close to the principal resonance, $%
\omega _{\mathrm{p}}\approx \omega _{0}$, the equivalent linear system is
characterized by the effective amplitude-dependent elasticity coefficient $%
k_{\mathrm{eff}}(v)$ and the effective damping factor $\lambda _{\mathrm{eff}%
}(v)$ (see chapter 7 in Ref.~\cite{Bogolyubov}):%
\begin{equation}
m\frac{d^{2}x}{dt^{2}}+\lambda _{\mathrm{eff}}(v)\frac{dx}{dt}+k_{\mathrm{eff%
}}(v)x=\epsilon F_{\mathrm{p}}\sin \omega _{\mathrm{p}}t,  \label{linearized}
\end{equation}%
\begin{eqnarray}
x &=&v\cos (\omega _{\mathrm{p}}t+\delta ),  \label{x} \\
k_{\mathrm{eff}}(v) &=&k_{0}-\frac{\epsilon }{\pi v}\int\limits_{0}^{2\pi }%
\widetilde{F}_{\mathrm{q}}(v,\psi )\cos \psi d\psi \equiv k_{0}+k_{\mathrm{q}%
},  \label{k_eff} \\
\lambda _{\mathrm{eff}}(v) &=&\lambda _{0}+\frac{\epsilon }{\pi v\omega _{0}}%
\int\limits_{0}^{2\pi }\widetilde{F}_{\mathrm{q}}(v,\psi )\sin \psi d\psi
\equiv \lambda _{0}+\lambda _{\mathrm{q}},  \label{lambda_eff}
\end{eqnarray}%
where $\widetilde{F}_{\mathrm{q}}(v,\psi )\equiv F_{\mathrm{q}}\left( x,%
\frac{dx}{dt}\right) =F_{\mathrm{q}}\left( v\cos \psi ,-\omega _{\mathrm{p}%
}v\sin \psi \right) $. Note that in Eq.~(\ref{linearized}) both $v$ and $%
\delta $ are time-dependent values.

Here, in equations (\ref{k_eff}) and (\ref{lambda_eff}), we have introduced
the \textit{parametric} elasticity coefficient $k_{\mathrm{q}}$ and damping
factor $\lambda _{\mathrm{q}}$. In this context the adjective \textit{quantum%
} is sometimes used instead of \textquotedblleft
parametric\textquotedblright\ to emphasize that it is the
qubit-state-dependent, i.e. it is defined by the quantum properties of the
coupled system. In what follows, by simply changing the notations we will
see that the parametric elasticity coefficient gives either parametric
inductance or parametric capacitance\textit{,} when coupling is inductive or
capacitive respectively, while the parametric damping factor will give us
the parametric resistance. Note that in equations (\ref{k_eff}) and (\ref%
{lambda_eff}) the parametric terms $k_{\mathrm{q}}$ and $\lambda _{\mathrm{q}%
}$ are of the first order in the small parameter of the problem $\epsilon $.

This linearization procedure allows to obtain important information even
without solving equations of motion. In particular, the effective resonance
frequency of the linearized system $\omega _{\mathrm{eff}}=\sqrt{k_{\mathrm{%
eff}}/m}$ gives the expression for the frequency shift%
\begin{equation}
\Delta \omega =\omega _{\mathrm{eff}}-\omega _{0}=\frac{k_{\mathrm{q}}}{%
2m\omega _{0}}.
\end{equation}

For physical interpretations it is important to emphasize that the
application of the linearization technique resulted in the substitution of
the nonlinear force by the linear one:%
\begin{equation}
\digamma \equiv \epsilon F_{\mathrm{q}}\left( x,\frac{dx}{dt}\right)
\longrightarrow \digamma _{\mathrm{q}}=-k_{\mathrm{q}}x-\lambda _{\mathrm{q}}%
\frac{dx}{dt}.  \label{Fq}
\end{equation}%
This latter \textquotedblleft parametric\textquotedblright\ force describes
the work done by the quantum system over the resonator; the respective
energy transfer during one period is the following%
\begin{equation}
W=\int\limits_{0}^{2\pi /\omega _{\mathrm{p}}}\digamma _{\mathrm{q}}\frac{dx%
}{dt}dt=-\pi \omega _{\mathrm{p}}v^{2}\lambda _{\mathrm{q}}.
\label{En_trasfer}
\end{equation}%
This, in dependence on the sign of the parametric damping factor $\lambda _{%
\mathrm{q}}$, describes periodical extraction or pumping of the energy by
the quantum system out of or into the resonator. This is known as the
\textit{Sisyphus damping and amplification} \cite{Grajcar08}.

The solution of equation (\ref{linearized}) in the first approximation in $%
\epsilon $ is given by the expression (\ref{x}) with the amplitude $v=v(t)$
and the phase shift $\delta =\delta (t)$\ slowly varying in time. For these
values the asymptotic expansion method gives the following system of
equations (see chapter 15 in Ref.~\cite{Bogolyubov})%
\begin{eqnarray}
\frac{dv}{dt} &=&-\frac{\lambda _{\mathrm{eff}}(v)}{2m}v-\frac{\epsilon F_{%
\mathrm{p}}}{m(\omega _{0}+\omega _{\mathrm{p}})}\cos \delta , \\
\frac{d\delta }{dt} &=&\omega _{\mathrm{eff}}(v)-\omega _{\mathrm{p}}+\frac{%
\epsilon F_{\mathrm{p}}}{mv(\omega _{0}+\omega _{\mathrm{p}})}\sin \delta .
\end{eqnarray}%
In the regime of stationary oscillations: $dv/dt=d\delta /dt=0$, and we
obtain equations for the amplitude $v$ and the phase shift $\delta $, which
can be written in the form%
\begin{eqnarray}
\tan \delta &=&\frac{k_{\mathrm{q}}(v)}{\omega _{0}\lambda _{\mathrm{eff}}(v)%
}, \\
v &=&-\frac{\epsilon F_{\mathrm{p}}\cos \delta }{\omega _{0}\lambda _{%
\mathrm{eff}}(v)}.
\end{eqnarray}%
In what follows it will be demonstrated that the phase shift $\delta $ and
the amplitude $v$ can be directly observed experimentally, which gives the
information about the quantum system through the values of the parametric
elasticity coefficient $k_{\mathrm{q}}$ and damping factor $\lambda _{%
\mathrm{q}}$.

\subsection{Inductive coupling with $LCR$ resonator. Parametric inductance}

\begin{figure}[t]
\includegraphics[width=7cm]{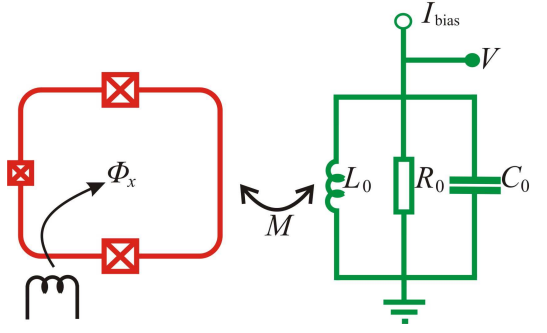}
\caption{(Color online) \textbf{Flux qubit coupled inductively to an }$LCR$%
\textbf{\ (tank) circuit.} The flux qubit is pierced by the magnetic flux $%
\Phi _{x}$ induced by the current in the controlling coil and by the current
in the tank's inductor. The qubit is coupled via the mutual inductance $M$
to the tank circuit. The resonant tank circuit consists of the inductor $%
L_{0}$, capacitor $C_{0}$, and resistor $R_{0}$; the circuit is biased with
an RF current $I_{\mathrm{bias}}$. The tank voltage $V$ is the measurable
value.}
\label{Fig:qb+LCR}
\end{figure}

Now we consider as an illustrative case the system of a flux qubit (with
geometrical inductance $L$ and average current $I_{\mathrm{qb}}$) coupled
inductively to the $LCR$ tank circuit, as shown in Fig.~\ref{Fig:qb+LCR}.
The approach, presented here, is the development of the theory in Refs.~\cite%
{Rifkin76, Greenberg02b, Smirnov03}. The quantum system is considered to be
weakly coupled via a mutual inductance $M$ to the classical tank circuit.
The circuit consists of the inductor $L_{0}$, capacitor $C_{0}$, and the
resistor $R_{0}$ connected, for the specification, in parallel. The tank
circuit is biased by the current $I_{\mathrm{bias}}$, and the voltage on it $%
V$ can be measured.

The flux qubit can be described by the pseudospin Hamiltonian \cite{Mooij99}%
\begin{eqnarray}
H &=&-\frac{\Delta }{2}\sigma _{x}-\frac{\varepsilon (t)}{2}\sigma _{z},
\label{H1qb} \\
\varepsilon (t) &=&\varepsilon _{0}+A\sin \omega t,  \label{e(t)}
\end{eqnarray}%
where the diagonal term $\varepsilon $ is the energy bias, the off-diagonal
term $\Delta $ is the tunneling amplitude between the wells (which
corresponds to the definite directions of the current in the loop) and $%
\sigma _{x,z}$ are the Pauli matrices.

To obtain the equation for the tank circuit voltage, we write down the
system of equations for the current in the three branches, namely, through
the inductor ($I_{\mathrm{L}}$), the capacitor ($I_{\mathrm{C}}$), and the
resistor ($I_{\mathrm{R}}$) (in particular, for systems with superconducting
elements see e.g. Ref.~\cite{Likharev}):%
\begin{eqnarray}
I_{\mathrm{bias}} &=&I_{\mathrm{L}}+I_{\mathrm{C}}+I_{\mathrm{R}},
\label{system} \\
I_{\mathrm{C}} &=&C_{0}\overset{\cdot }{V},\text{ }I_{\mathrm{R}}=V/R_{0}, \\
V &=&L_{0}\dot{I}_{\mathrm{L}}-\dot{\Phi}_{\mathrm{e}},  \label{4}
\end{eqnarray}%
where $\Phi _{\mathrm{e}}$ is the flux through the inductor of the tank
circuit. This flux is the response of the quantum system to the flux,
induced in it by the current $I_{\mathrm{L}}$. It follows that the voltage $%
V $ in the current-biased tank circuit ($I_{\mathrm{bias}}=I_{\mathrm{A}%
}\sin \omega _{\mathrm{p}}t$) is described by the following nonlinear
equation

\begin{equation}
C_{0}\frac{d^{2}V}{dt^{2}}+R_{0}^{-1}\frac{dV}{dt}+L_{0}^{-1}V=-\frac{\dot{%
\Phi}_{\mathrm{e}}(V,\dot{V})}{L_{0}}+I_{\mathrm{A}}\omega _{\mathrm{p}}\cos
\omega _{\mathrm{p}}t.  \label{eq_for_V}
\end{equation}%
The external flux $\Phi _{\mathrm{e}}$ is assumed to be proportional to the
coupling parameter $k^{2}=M^{2}/LL_{0}\ll 1$ and to depend on time via the
voltage $V$ and its time derivative $\dot{V}$. Equation~(\ref{eq_for_V}) for
the voltage $V$ coincides with the nonlinear equation~(\ref{eq4x}) for the
variable $x$ with obvious change of the notations.

Thus, the formalism presented in the previous subsection is directly
applicable for the given problem. Specifically, in the first order
approximation with respect to the coupling parameter $k^{2}$ and close to
the principal resonance ($\omega _{\mathrm{p}}\approx \omega _{0}\equiv 1/%
\sqrt{L_{0}C_{0}}$), the equivalent linear system is characterized by the
effective resistance $R_{\mathrm{eff}}$ and inductance $L_{\mathrm{eff}}$ as
following

\begin{equation}
C_{0}\frac{d^{2}V}{dt^{2}}+R_{\mathrm{eff}}^{-1}\frac{dV}{dt}+L_{\mathrm{eff}%
}^{-1}V=I_{A}\omega _{\mathrm{p}}\cos \omega _{\mathrm{p}}t,  \label{eff}
\end{equation}

\begin{eqnarray}
V &=&v\cos (\omega _{\mathrm{p}}t+\delta ),  \label{V} \\
\frac{1}{R_{\mathrm{eff}}(v)} &=&\frac{1}{R_{0}}+\frac{1}{R_{\mathrm{q}}(v)},
\label{Reff} \\
\frac{1}{L_{\mathrm{eff}}(v)} &=&\frac{1}{L_{0}}+\frac{1}{L_{\mathrm{q}}(v)}.
\label{Leff}
\end{eqnarray}%
Here $Q_{0}=\omega _{0}C_{0}R_{0}$ is the quality factor of the unloaded
tank circuit (at $\Phi _{\mathrm{e}}=0$) and the \textit{parametric }%
(qubit's-state dependent)\textit{\ resistance }$R_{\mathrm{q}}$\textit{\ }and%
\textit{\ inductance }$L_{\mathrm{q}}$ are given by the formulas%
\begin{eqnarray}
\frac{1}{R_{\mathrm{q}}(v)} &=&-\frac{Q_{0}}{\pi vR_{0}}\int\limits_{0}^{2%
\pi }\widetilde{\dot{\Phi}}_{\mathrm{e}}(v,\psi )\sin \psi d\psi ,
\label{Rq} \\
\frac{1}{L_{\mathrm{q}}(v)} &=&\frac{1}{\pi vL_{0}}\int\limits_{0}^{2\pi }%
\widetilde{\dot{\Phi}}_{\mathrm{e}}(v,\psi )\cos \psi d\psi ,  \label{Lq}
\end{eqnarray}%
where $\widetilde{\dot{\Phi}}_{\mathrm{e}}(v,\psi )\equiv \dot{\Phi}_{%
\mathrm{e}}(V,\dot{V})=\dot{\Phi}_{\mathrm{e}}(v\cos \psi ,-v\omega _{%
\mathrm{p}}\sin \psi )$. The resonant frequency $\omega _{\mathrm{eff}}$
becomes amplitude-dependent and is shifted by

\begin{equation}
\Delta \omega =\omega _{\mathrm{eff}}(v)-\omega _{0}=\frac{\omega _{0}L_{0}}{%
2L_{\mathrm{q}}(v)}.  \label{weff}
\end{equation}%
The phase shift $\delta $ and the amplitude $v$ depend on the probing
frequency detuning $\xi _{0}\equiv \frac{\omega _{0}-\omega _{\mathrm{p}}}{%
\omega _{0}}$ and the qubit state (via $L_{\mathrm{q}}$ and $R_{\mathrm{q}}$%
). In the stationary regime they are given by the solution of the system of
equations

\begin{equation}
\left\{
\begin{array}{c}
\tan \delta =2Q_{0}\frac{R_{\mathrm{eff}}}{R_{0}}\left( \xi _{0}+\frac{%
L_{0}-L_{\mathrm{eff}}}{2L_{0}}\right) , \\
v=I_{A}R_{\mathrm{eff}}\cos \delta ,%
\end{array}%
\right.  \label{with_L_R_eff}
\end{equation}%
which can also be rewritten alternatively in terms of the effective quality
factor $Q_{\mathrm{eff}}=\omega _{0}C_{0}R_{\mathrm{eff}}(v)$ and effective
frequency shift $\xi _{\mathrm{eff}}=\left[ \omega _{\mathrm{eff}}(v)-\omega
_{\mathrm{p}}\right] /\omega _{0}$.

Thus, the observable values -- the amplitude $v$ and the phase shift $\delta
$ -- are defined by equations (\ref{with_L_R_eff}), which depend on the
response of the measurable system, $\Phi _{e}(V,\dot{V})$. As we discussed
above, strictly speaking, the dynamics of the tank circuit has to be
considered jointly with the dynamics of the qubit (corresponding
calculations see e. g. in \cite{Greenberg08}). However, in what follows we
consider two illustrative limiting cases, when the dynamics of the qubit can
be treated separately from the dynamics of the tank circuit. For
simplification we introduce phenomenologically the relaxation time $T_{1}$
which is caused by the coupling to the environment and to the tank as well.

\textit{1. Low-quality\ qubit (}$T_{1}\ll T$\textit{): phase shift probes
the parametric inductance of qubit.}

First case which allows to detach the equations for the qubit and resonator,
is when all the qubit's characteristic times, and in particular the
relaxation time $T_{1}$, are smaller than the tank's period $T=2\pi /\omega
_{0}$. Then the equations for the tank voltage can be averaged over the
period of fast oscillations. Then the time derivative of the flux $\Phi _{e}$%
, induced by the qubit in the tank circuit can be described as
\begin{equation}
\dot{\Phi}_{e}=M\dot{I}_{\mathrm{qb}}=M\frac{\partial I_{\mathrm{qb}}}{%
\partial \Phi }\dot{\Phi},  \label{dFie}
\end{equation}%
where $\Phi =\Phi _{\mathrm{dc}}+MI_{\mathrm{L}}$ is the flux in the qubit's
loop, which consists of the time-independent part $\Phi _{\mathrm{dc}}$ and
of the flux, induced by the current $I_{\mathrm{L}}$ in the tank's inductor.
This can be rewritten by introducing the effective inductance of the qubit, $%
\mathcal{L}^{-1}=\partial I_{\mathrm{qb}}(\Phi )/\partial \Phi $, and the
characteristic inductance value $\widetilde{L}=M^{2}\mathcal{L}^{-1}$. Then $%
\dot{\Phi}_{e}=\widetilde{L}(I_{\mathrm{L}})\dot{I}_{\mathrm{L}}$ and for
the tank voltage we have $V=L_{0}\dot{I}_{\mathrm{L}}-\dot{\Phi}_{e}=(L_{0}-%
\widetilde{L}(I_{\mathrm{L}}))\dot{I}_{\mathrm{L}}$. In the first
approximation in $k^{2}$ in the expression $\dot{\Phi}_{e}=\widetilde{L}(I_{%
\mathrm{L}})\dot{I}_{\mathrm{L}}$ we can insert $I_{\mathrm{L}}$ found from
this equation

\begin{equation}
I_{\mathrm{L}}(t)\approx \frac{1}{L_{0}}\int Vdt\approx \frac{v}{\omega
_{0}L_{0}}\sin (\omega _{\mathrm{p}}t+\delta ).  \label{I_L}
\end{equation}%
Then from Eqs.~(\ref{Rq}-\ref{Lq}) we have $R_{\mathrm{q}}^{-1}=0$ (hence $%
R_{\mathrm{eff}}=R_{0}$) and
\begin{equation}
\frac{L_{0}}{L_{\mathrm{q}}}=\frac{k^{2}L}{\pi }\int\limits_{0}^{2\pi }%
\mathcal{L}^{-1}(v,\psi )\cos ^{2}\psi d\psi ,  \label{beta_for_quick}
\end{equation}%
where the qubit's effective inductance is defined by the total flux $\Phi $,
piercing the qubit's loop%
\begin{equation}
\mathcal{L}^{-1}(v,\psi )\equiv \left. \frac{\partial I_{\mathrm{qb}}(\Phi )%
}{\partial \Phi }\right\vert _{\Phi =\Phi _{\mathrm{dc}}+\frac{M}{%
L_{0}\omega _{0}}v\sin \psi }.  \label{Jos_induct}
\end{equation}%
Then for the phase shift $\delta $ and the voltage amplitude $v$\ we obtain
\cite{Shevchenko08}%
\begin{equation}
\tan \delta \approx 2Q_{0}\xi _{0}+Q_{0}\frac{L_{0}}{L_{\mathrm{q}}},\text{
\ }v\approx I_{A}R_{0}\cos \delta ,  \label{a_and_v}
\end{equation}%
which is the generalization of the result of Ref.~\cite{Greenberg02a} for
the case when the qubit can be in the superpositional state, which is taken
into account here by the expectation value of the current $I_{\mathrm{qb}}$.
If the bias current amplitude $I_{A}$ is small enough to be ignored in Eq.~(%
\ref{Jos_induct}), then%
\begin{eqnarray}
L_{\mathrm{q}}^{-1} &=&k^{2}\frac{L}{L_{0}}\mathcal{L}^{-1},\text{ \ \ \ \ \
}\mathcal{L}^{-1}\approx \frac{\partial I_{\mathrm{qb}}}{\partial \Phi _{%
\mathrm{dc}}},  \notag \\
\tan \delta &\approx &2Q_{0}\xi _{0}+k^{2}Q_{0}L\mathcal{L}^{-1},\text{ \ }%
v\approx I_{A}R_{0}\cos \delta .  \label{QIMT}
\end{eqnarray}%
At the resonant frequency $\xi _{0}=0$, the phase shift $\delta $\ is
proportional to the inverse inductance of the qubit $\mathcal{L}^{-1}$. Here
it is worthwhile to emphasize the expression for the parametric inductance,
which is expressed via the derivative of the expectation value of the
current in the qubit's loop $I_{\mathrm{qb}}=-I_{\mathrm{p}}\left\langle
\sigma _{z}\right\rangle $,%
\begin{equation}
L_{\mathrm{q}}^{-1}=-L_{0}^{-1}k^{2}\frac{LI_{\mathrm{p}}}{\Phi _{0}}\frac{%
\partial \left\langle \sigma _{z}\right\rangle }{\partial f_{\mathrm{dc}}}.
\label{Lq2}
\end{equation}

\textit{2. Higher-quality\ qubit (}$T_{1}\lesssim T$\textit{): parametric
resistance due to qubit's lagging.}

Another illustrative situation, where the qubit's dynamics can be considered
separately from the resonator's one, is the case when the qubit relaxation
time $T_{1}$ is of the same order as the tank's period $T$, namely, $%
T_{1}\lesssim T$. The qubit's response to the resonator probing signal can
be phenomenologically described by introducing the lagging time $t^{\prime
}=t-T_{1}$, so that instead of Eq. (\ref{dFie}) we have%
\begin{equation}
\dot{\Phi}_{e}(t)=\widetilde{L}(I_{\mathrm{L}}(t^{\prime }))\dot{I}_{\mathrm{%
L}}(t^{\prime }).  \label{Fi_e_}
\end{equation}%
In this way, the qubit's response depends on the current in the tank $I_{%
\mathrm{L}}=I_{\mathrm{L}}(t^{\prime })$, which is given by
\begin{equation}
I_{\mathrm{L}}(t^{\prime })\approx \frac{v}{\omega _{0}L_{0}}\left( C\sin
(\omega _{\mathrm{p}}t+\delta )-S\cos (\omega _{\mathrm{p}}t+\delta )\right)
\label{prime}
\end{equation}%
with $S=\sin (\omega _{\mathrm{p}}T_{1})$ and $C=\cos (\omega _{\mathrm{p}%
}T_{1})$. For the small bias current Eqs.~(\ref{Reff}-\ref{Leff}) and (\ref%
{Fi_e_}-\ref{prime}) result in the following expressions for the parametric
inductance and resistance
\begin{align}
L_{0}/L_{\mathrm{q}}& \approx C\cdot k^{2}L\mathcal{L}^{-1},  \notag \\
R_{0}/R_{\mathrm{q}}& \approx -S\cdot k^{2}Q_{0}L\mathcal{L}^{-1}.
\label{Lq&Rq}
\end{align}%
By analogy with Eq.~(\ref{Lq2}), the latter phenomenological equation can be
rewritten in the form explicitly demonstrating its quantum character,%
\begin{equation}
R_{\mathrm{q}}^{-1}=S\frac{k^{2}Q_{0}}{R_{0}}\frac{LI_{\mathrm{p}}}{\Phi _{0}%
}\frac{\partial \left\langle \sigma _{z}\right\rangle }{\partial f_{\mathrm{%
dc}}}.  \label{Rq2}
\end{equation}%
By making use of Eq.~(\ref{En_trasfer}), we obtain that the energy
transferred from qubit into the resonator (or, out of the resonator, for the
opposite sign) during one period is%
\begin{equation}
W=-\pi \omega _{\mathrm{p}}v^{2}R_{\mathrm{q}}^{-1}.  \label{En_trasfer2}
\end{equation}

We emphasize here that both the parametric inductance and resistance in Eq.~(%
\ref{Lq&Rq}) are proportional to the qubit's inductance $\mathcal{L}$. Then,
one obtains equations for $\delta $ and $v$, which are simplified in the
first approximation in $k^{2}Q_{0}L\mathcal{L}^{-1}$. In this case for the
probing frequency equal to the resonant one, $\xi _{0}=0$, the resulting
formulas are%
\begin{align}
\tan \delta & \approx C\cdot k^{2}Q_{0}L\mathcal{L}^{-1},  \label{alpha&v} \\
\frac{v}{I_{A}R_{0}}& \approx 1+S\cdot k^{2}Q_{0}L\mathcal{L}^{-1}.  \notag
\end{align}%
Note that both the phase shift and amplitude are related to the qubit's
effective inductance $\mathcal{L}$, which explains their similar behavior in
experiment. These equations are useful for the analysis of the experimental
results, as it will be demonstrated in Section 4.

\subsection{Capacitive coupling with nanomechanical resonator. Parametric
capacitance}

Consider now the charge qubit capacitively coupled to a resonator. In this
case, like in the one considered above, the resonator can be the tank
circuit. Alternatively, the resonator can be a nanomechanical resonator
(NR), as in Ref.~\cite{LaHaye09}. For the illustrative purpose, we consider
here this latter case.

\begin{figure}[t]
\includegraphics[width=8 cm]{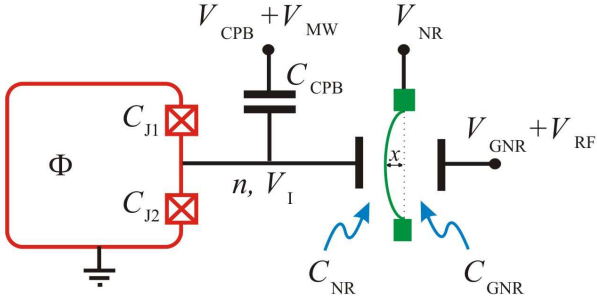}
\caption{(Color online) \textbf{Charge qubit probed by a nanomechanical
resonator.} The charge qubit is the Cooper-pair box, controlled by the
magnetic flux $\Phi $ and the gate voltage $V_{\mathrm{CPB}}+V_{\mathrm{MW}}$%
. The resonator probing the qubit's state here is the NR, which is
characterized by the displacement at the midpoint $x$. The voltage-biased NR
is measured through its resonance frequency shift $\Delta \protect\omega _{%
\mathrm{NR}}$. \protect\cite{Shevchenko11}}
\label{Fig:qb+NR}
\end{figure}

The split-junction charge qubit (shown in red in Fig.~\ref{Fig:qb+NR})
consists of a small island between two Josephson junctions (also called
Cooper-pair box), whose state is controlled by the magnetic flux $\Phi $ and
the gate voltage $V_{\mathrm{CPB}}+V_{\mathrm{MW}}$. Here $V_{\mathrm{CPB}}$
is the dc voltage used to tune the energy levels of the qubit and $V_{%
\mathrm{MW}}=V_{\mu }\sin \omega t$ is the microwave signal used to change
the energy-level occupations. The driven Cooper-pair box is described in the
two-level approximation by the Hamiltonian in the \textquotedblleft
charge\textquotedblright\ representation, Eqs.~(\ref{H1qb}-\ref{e(t)}),
where the tunnel splitting $\Delta $ is equal to the Josephson energy
controlled by the magnetic flux $\Phi $: $\Delta =E_{\mathrm{J}0}\left\vert
\cos (\pi \Phi /\Phi _{0})\right\vert $. The charging energy and the driving
amplitude are the following $\varepsilon _{0}=-8E_{\mathrm{C}}(n_{\mathrm{g}%
}-1/2)$ and $A=-8E_{\mathrm{C}}n_{\mathrm{\mu }}$, where the Coulomb energy $%
E_{\mathrm{C}}=e^{2}/2C_{\Sigma }$ is defined by the total capacitance $%
C_{\Sigma }=2C_{\mathrm{J}}+C_{\mathrm{CPB}}+C_{\mathrm{NR}}$ and the
effective Josephson capacitance is introduced $2C_{\mathrm{J}}\equiv C_{%
\mathrm{J}1}+C_{\mathrm{J}2}$, the dimensionless driving amplitude $n_{\mu
}=C_{\mathrm{CPB}}V_{\mu }/2e$. The dimensionless polarization charge $n_{%
\mathrm{g}}=n_{\mathrm{NR}}+n_{\mathrm{CPB}}$ is the fractional part of the
respective polarization charges in two capacitances: $n_{\mathrm{NR}%
}=\left\{ C_{\mathrm{NR}}V_{\mathrm{NR}}/2e\right\} $ and $n_{\mathrm{CPB}%
}=\left\{ C_{\mathrm{CPB}}V_{\mathrm{CPB}}/2e\right\} $.

The Cooper-pair box here is formed by four capacitors, $C_{\mathrm{J}1}$, $%
C_{\mathrm{J}2}$, $C_{\mathrm{CPB}}$, and $C_{\mathrm{NR}}$. One of the
plates of the latter capacitor is formed by the NR. The displacement of the
NR $x$ is much smaller than the distance $d$ between the plates. Then the
capacitance between the NR and the qubit reads

\begin{equation}
C_{\mathrm{NR}}(x)\approx C_{\mathrm{NR}}+\frac{\partial C_{\mathrm{NR}}}{%
\partial x}x\equiv C_{\mathrm{NR}}\left( 1+\frac{x}{\xi }\right) .
\label{CNR(x)}
\end{equation}%
Here $C_{\mathrm{NR}}$ stands for the capacitance value at the zero
displacement. The displacement of the NR influences the qubit through the
changes in the polarization charge; to make it significant, a large dc
voltage $V_{\mathrm{NR}}$ is applied. On the other side, the NR is biased by
dc and rf voltages $V_{\mathrm{GNR}}$ and $V_{\mathrm{RF}}$ through the
capacitance $C_{\mathrm{GNR}}$.

One of the approaches to describe the system qubit-resonator is to introduce
the parametric capacitance as following (for more details see Ref.~\cite%
{Shevchenko11}). Let us introduce the effective capacitance, as it is
demonstrated in Fig.~\ref{Fig:effectiveL&C}(c), by differentiating the
charge $Q_{\mathrm{NR}}$ of the capacitor $C_{\mathrm{NR}}$ \cite%
{Sillanpaa05, Duty05, Johansson06}: $C_{\mathrm{eff}}=\partial Q_{\mathrm{NR}%
}/\partial V_{\mathrm{NR}}$. Then, for the charge $Q_{\mathrm{NR}}=(V_{%
\mathrm{NR}}-V_{\mathrm{I}})C_{\mathrm{NR}}$ with the island's voltage given
by $V_{\mathrm{I}}=2e(n_{\mathrm{g}}-\left\langle n\right\rangle )/C_{\Sigma
}$, we obtain $C_{\mathrm{eff}}=C_{\mathrm{geom}}+C_{\mathrm{q}}$, which
consists of the \textit{parametric capacitance}%
\begin{equation}
\text{ }C_{\mathrm{q}}=\frac{C_{\mathrm{NR}}^{2}}{C_{\Sigma }}\frac{\partial
\left\langle n\right\rangle }{\partial n_{\mathrm{g}}}
\end{equation}%
and the geometric capacitance $C_{\mathrm{geom}}$\
\begin{equation}
C_{\mathrm{geom}}=\frac{C_{\mathrm{NR}}(C_{\Sigma }-C_{\mathrm{NR}})}{%
C_{\Sigma }}\approx \frac{2C_{\mathrm{J}}C_{\mathrm{NR}}}{2C_{\mathrm{J}}+C_{%
\mathrm{NR}}}\approx C_{\mathrm{NR}},\text{ }  \label{Cgeom}
\end{equation}%
where the approximations are valid for $C_{\mathrm{CPB}}\ll C_{\mathrm{J}%
},C_{\mathrm{NR}}$ and $C_{\mathrm{NR}}\ll C_{\mathrm{J}}$ respectively.
Then one can consider the force $F_{\mathrm{NR}}$, which acts on the NR from
the left electrode, as the electrostatic force from the effective
capacitance [see Fig.~\ref{Fig:effectiveL&C}(c)]: $F_{\mathrm{NR}}=\frac{1}{2%
}\frac{\partial }{\partial x}\left( C_{\mathrm{eff}}V_{\mathrm{NR}%
}^{2}\right) $. Then the term with the parametric capacitance, in which $C_{%
\mathrm{NR}}^{2}\approx C_{\mathrm{NR}}^{2}\left( 1+x/\xi \right) ^{2}$,
results in the following resonance frequency shift of the NR%
\begin{eqnarray}
\frac{\Delta \omega _{\mathrm{NR}}}{\omega _{\mathrm{NR}}} &=&-\frac{\beta
C_{\mathrm{\Sigma }}}{C_{\mathrm{NR}}^{2}}C_{\mathrm{q}}=-\beta \frac{%
\partial \left\langle n\right\rangle }{\partial n_{\mathrm{g}}}=-\frac{\beta
}{2}\frac{\partial \left\langle \sigma _{z}\right\rangle }{\partial n_{%
\mathrm{g}}},  \label{DwNR_2} \\
\beta &=&\frac{1}{m\omega _{\mathrm{NR}}^{2}C_{\Sigma }}\left( \frac{C_{%
\mathrm{NR}}V_{\mathrm{NR}}}{\xi }\right) ^{2}.  \notag
\end{eqnarray}

We would like to note that the results obtained for the system qubit-NR can
be definitely extended to other systems. For example the charge qubit can be
coupled to a tank circuit instead of a NR. In contrast to the inductive
coupling, considered in the previous subsection, here we mean capacitive
coupling. Then it is straightforward to obtain the expression for the
measurable value, the tank circuit phase shift at resonance frequency, $\xi
_{0}=0$, \cite{Shevchenko11}%
\begin{equation}
\tan \delta \approx Q_{0}\frac{C_{\mathrm{q}}}{C_{0}},
\end{equation}%
cf. Eq.~(\ref{a_and_v}), where the phase shift probes the parametric
inductance. In section 4 it will be demonstrated how these expressions can
be used for the description of the realistic system.

\section{3. Dynamical behavior of a two-level system}

Application of the semiclassical theory, presented in the previous
subsection, to the description of the qubits-resonator system makes possible
to separate the slow dynamics of the resonator from the fast dynamics of the
qubits system. This allows to consider first the dynamics of a qubit or a
system of qubits. Then, the resonator can monitor the state of the system of
qubits. In this section we will outline the description of the multiphoton
processes in a qubit, while the presentation of the specific results is the
subject of the next two sections.

Initialization and manipulation of the qubit's systems require certain
external signals. The principal features of the driven system are captured
for the harmonic driving, Eq. (\ref{e(t)}), to which we limit our
consideration. Different theoretical approaches can be used for a driven
two-level system, which is described in the books and reviews \cite{Blum,
Weiss, Faisal, Nakamura02, Grifoni98, Chu04}. The choice of the formalism
depends on the formulation of a problem and on the parameters of the system,
such as the bias offset $\varepsilon _{0}$, driving amplitude $A$ and
frequency $\omega $. The clear description can be given for the temporal
dynamics in the so-called adiabatic-impulse model, where the driven
evolution is considered adiabatic far from the avoided-level crossings with
the impulse-type Landau-Zener transitions, when the energy distance is
minimal \cite{Garraway97, Forre04, Shevchenko10}. As the result of this
theory, the overall dynamics is described by the long-time Rabi-type
oscillations of the level occupation probabilities with the step-like
features due to the Landau-Zener transitions.

Another technique, which can be more convenient for the resonant driving, is
the rotating wave approximation (RWA) \cite{Son09, Wen09, Ferron10}. It
consists in neglecting the rapidly oscillating (non-resonant) terms. The
common approach for making use of this approximation is taking small driving
amplitudes, $A\ll \Delta E$. Then, the first-order consideration gives usual
Rabi oscillations of the level occupation probabilities close to the
position of the one-photon resonance, where $\omega \approx \Delta E/\hbar $%
. In the $k$-th approximation, the resonant excitation appears close to the
parameters, where the energy of $k$ photons matches the qubit's energy
distance \cite{Coh-Tan, Delone}%
\begin{equation}
k\hbar \omega =\Delta E.  \label{multi}
\end{equation}%
The time evolution is described by the multiphoton Rabi oscillations \cite%
{Shirley65}, while the time-averaged upper-level occupation probability has
the Lorentzian shape with the maximum equal to $1/2$ at the exact resonance
defined by Eq.~(\ref{multi}).

With increasing the driving amplitude the resonances shift \cite{Krainov80}
from their positions given by the perturbation theory and defined by the
exact multiphoton relation (\ref{multi}). The first-order correction to the
position of the resonances is the so-called Bloch-Siegert shift \cite%
{Coh-Tan}; it was demonstrated for the superconducting qubits in Ref. \cite%
{Tuorila10}. Thus, in general, the position of the multiphoton resonances is
amplitude-dependent.

For the description of the strongly driven qubits, another formulation of
the RWA can be used. There, the minimal energy level splitting $\Delta $ is
the small parameter, namely, it is assumed $\Delta \ll \sqrt{A\omega }$ \cite%
{LopezCastillo92, Oliver05, Ashhab07}. Then the $k$-photon excitation
appears close to the resonant parameters, given by the relation $\varepsilon
_{0}=k\hbar \omega $. There, the upper-level occupation probability $P_{%
\mathrm{up}}(t)$\ oscillates with the frequency $\Omega _{\mathrm{R}}=\sqrt{%
(\varepsilon _{0}-k\hbar \omega )^{2}+\Delta _{k}^{2}}$ with the
renormalized splitting $\Delta _{k}=\Delta J_{k}\left( A/\omega \right) $; $%
J_{k}$\ is the Bessel function. The time-averaged probability in the
vicinity of the $k$-th resonance is given by%
\begin{equation}
\overline{P}_{\mathrm{up}}=\frac{1}{2}\frac{\Delta _{k}^{2}}{(\varepsilon
_{0}-k\hbar \omega )^{2}+\Delta _{k}^{2}}.
\end{equation}

Being time averaged, the Rabi oscillations are described by the Lorentzian
dependence of the upper-level occupation on the system's parameters (the
bias or the driving frequency) \cite{Tornes08}. Here arises an interesting
and important problem of distinction of the respective quantum oscillations
from their classical counterparts, which are the parametric resonances. This
was the subject of Refs.~\cite{Gronbech05, Marchese07, ShevchenkoOZ08}.

The most straightforward approach for the numerical description of the
dynamics of a two-level system is the solution of the Schr\"{o}dinger
equation~\cite{Shevchenko05}. Then, the influence of the dissipation can be
taken into account phenomenologically by introducing energy and phase
relaxation times, $T_{1}$ and $T_{2}$, and solving the respective Bloch
equation~\cite{Blum}. Instead, in the more general approach, the dissipative
environment can be described as an ensemble of oscillators, which would
result in the Bloch-Redfield equation for the reduced density matrix~\cite%
{Leggett87, Goorden05}. This latter formalism will be demonstrated in Sec.~5
being applied to the specific case of the two-qubit system.

Note that the multiphoton transitions can also be driven by the bichromatic
field, when the energy level distance $\Delta E$ is matched by the energy of
several photons of one (say, microwave-) frequency plus several photons of
another (say, radio-) frequency. Such transitions were studied both in
microscopic systems \cite{Delone, Saiko07}, and in the Josephson-junction
qubits \cite{Saito06, Gunnarsson08, Paila09}. Also for the case of a flux
qubit it was demonstrated that the persistence of Rabi oscillations can be
supported by either the low-frequency signal \cite{Greenberg05} or induced
by noise \cite{Omelyanchouk09}.

\section{4. Excitation of a superconducting qubit}

Let us get back to the qubit-resonator systems. In the previous Section we
have discussed a modification of the qubit states (and therefore its
observables) under different types of excitations. A natural next step is to
analyze the corresponding (via qubits) change of the resonator properties.
In this section we demonstrate this by presenting respective theoretical
results for different realizations of the qubit-resonator systems, making
use of the theory presented in the previous two sections. The emphasis is
made on demonstrating the consistency of the theoretical results with the
experimental ones.

\subsection{Inductance of superconducting qubits}

Consider a qubit biased with a DC flux $\Phi _{\mathrm{dc}}$ and driven with
an AC flux $\Phi _{\mathrm{ac}}\sin \omega t$, introducing $f_{\mathrm{dc}%
}=\Phi _{\mathrm{dc}}/\Phi _{0}-1/2$ and $f_{\mathrm{ac}}=\Phi _{\mathrm{ac}%
}/\Phi _{0}.$ In order to get the effective inductance $\mathcal{L}$, as
defined by Eq.~(\ref{Jos_induct}), we have to calculate the average current
in the qubit: $I_{\mathrm{qb}}=\left\langle I\right\rangle =\mathrm{Tr}%
\left( \rho I\right) $, where $I=I_{\mathrm{p}}\sigma _{z}$ is the current
operator defined with the amplitude $I_{\mathrm{p}}$ and the Pauli matrix $%
\sigma _{z}$. We calculate the reduced density matrix $\rho $ with the Bloch
equations \cite{Blum, Shevchenko05} which include phenomenological
relaxation times, $T_{1}$ and $T_{2}$. It is convenient to express the
density matrix in the energy representation: $\rho =\left( 1/2\right) \left(
\tau _{0}+X\tau _{x}+Y\tau _{y}+Z\tau _{z}\right) $, where $\tau _{i}$ are
the Pauli matrices for this basis and $\tau _{0}$ stands for the unity
matrix. The value $Z=\left\langle \widehat{\tau }_{z}\right\rangle $ is
equal to the difference between the populations of the ground and excited
states.

Let us find now the explicit expressions for the effective qubit's
inductance for both the interferometer-type (split-junction) charge qubit
\cite{Krech02, Zorin02} and flux qubit \cite{Mooij99}. For the
interferometer-type charge qubit, as considered in detail in Ref.~\cite%
{Shnyrkov06}, the circulating current $I_{0}$ is flux-dependent and Eqs.~(%
\ref{QIMT}) show that there are two terms contributing in the tank circuit's
phase shift,%
\begin{equation}
\tan \delta \approx \frac{k^{2}QL}{\Phi _{0}}\left( \frac{\partial I_{0}}{%
\partial f_{\mathrm{dc}}}Z+I_{0}\frac{\partial Z}{\partial f_{\mathrm{dc}}}%
\right) .  \label{ch_qb}
\end{equation}%
In a classical system (where the current has a definite direction) or in the
ground state, the difference between the energy level's populations is
constant, $Z=const$, and the second term in Eq.~(\ref{ch_qb}) is zero. In
contrast, for the quantum system the interplay between these two terms is
essential. At this point it is worthwhile to notice that the second term can
dominate at resonant excitation, as it was the case in the work \cite%
{Shnyrkov06} (see also below). This means that the second (\textquotedblleft
quantum\textquotedblright ) term can significantly increase the sensitivity
of the impedance measurement technique, as compared to the classical
situation described by the first term in Eq.~(\ref{ch_qb}).

Consider now the case of a flux qubit. The current operator is defined in
the flux basis \cite{Mooij99}, $I=I_{\mathrm{p}}\sigma _{z}$, where $I_{%
\mathrm{p}}$ stands for the amplitude value of the persistent current, and
hence the value $\left\langle \sigma _{z}\right\rangle $ defines the
difference between the probabilities of the clockwise and counter-clockwise
current directions in the loop: $\left\langle \sigma _{z}\right\rangle
=P_{\downarrow }-P_{\uparrow }=2P_{\downarrow }-1$. Then with Eqs.~(\ref%
{QIMT}) we obtain%
\begin{equation}
\tan \delta \approx k^{2}Q\frac{LI_{\mathrm{p}}}{\Phi _{0}}2\frac{\partial
P_{\downarrow }}{\partial f_{\mathrm{dc}}}.  \label{flux_qb1}
\end{equation}%
In the energy representation we rewrite Eq.~(\ref{flux_qb1})%
\begin{equation}
\tan \delta \approx k^{2}Q\frac{LI_{\mathrm{p}}}{\Phi _{0}}\frac{\partial }{%
\partial f_{\mathrm{dc}}}\left( \frac{\Delta }{\Delta E}X-\frac{I_{\mathrm{p}%
}\Phi _{0}f_{\mathrm{dc}}}{\Delta E}Z\right) .  \label{tan(a)_full}
\end{equation}%
Here $\Delta E=\sqrt{\Delta ^{2}+(I_{\mathrm{p}}\Phi _{0}f_{\mathrm{dc}})^{2}%
}$ is the distance between the stationary energy levels.

After the time-averaging over the driving period $2\pi /\omega $, this
expression is written as following

\begin{equation}
\tan \delta \approx -k^{2}Q\frac{LI_{\mathrm{p}}^{2}}{\Delta }\left( \frac{%
\Delta ^{3}}{\Delta E^{3}}+\frac{\Delta }{\Delta E}f_{\mathrm{dc}}\frac{%
\partial }{\partial f_{\mathrm{dc}}}\right) Z.  \label{approx}
\end{equation}%
If a qubit is resonantly excited with the driving frequency $\omega $, then
the partial energy levels occupation probability $Z$ has the
Lorentzian-shape dependence on $f_{\mathrm{dc}}$. It follows that the
derivative $\partial Z/\partial f_{\mathrm{dc}}$ takes the shape of a
hyperbolic-like structure, i.e. it changes from a peak to a dip in the point
of the resonance at $\Delta E(f_{\mathrm{dc}})\approx k\hbar \omega $.

\subsection{Equilibrium-state measurement}

\begin{figure}[t]
\includegraphics[width=8cm]{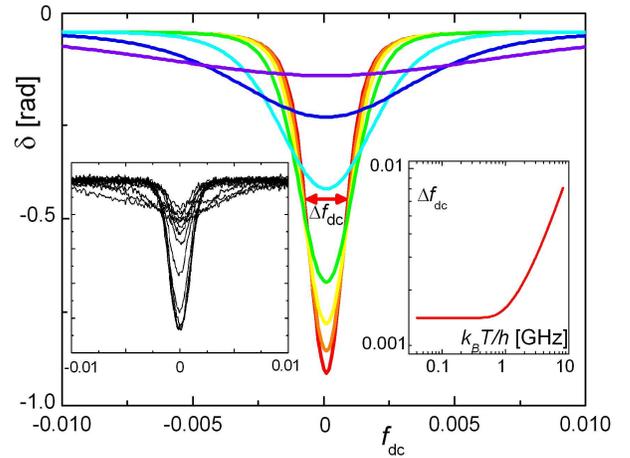}
\caption{(Color online) \textbf{The equilibrium-state measurement}. The
dependence of the tank phase shift on the flux detuning $f_{\mathrm{dc}%
}=\Phi _{\mathrm{dc}}/\Phi _{0}-1/2$, when the qubit is thermally excited.
The curves are plotted for $k_{\mathrm{B}}T/h=0.2$, $0.5$, $0.7$, $1$, $2$, $%
4$, and $8$ GHz. Left inset: corresponding experimental results \protect\cite%
{Grajcar04}. Right inset: temperature dependence of the width $\Delta f_{%
\mathrm{dc}}$ of the dip at half-depth in the phase shift, shown in the main
panel. \protect\cite{ShevchenkoPG08}}
\label{Fig:gr_st_vs_T}
\end{figure}

For the description of the measurement of a \textit{flux qubit in the
thermal equilibrium} one has to put $X=0$ and $Z=\tanh \left( \Delta E/2k_{%
\mathrm{B}}T\right) $ in Eq.~(\ref{tan(a)_full}),

\begin{equation}
\tan \delta \approx -k^{2}Q\frac{LI_{\mathrm{p}}^{2}}{\Delta }\left( \frac{%
\Delta ^{3}}{\Delta E^{3}}+\frac{\Delta }{\Delta E}f_{\mathrm{dc}}\frac{%
\partial }{\partial f_{\mathrm{dc}}}\right) \tanh \left( \frac{\Delta E}{2k_{%
\mathrm{B}}T}\right) .  \label{alfagr}
\end{equation}%
The\textit{\ ground-state measurement} at $k_{\mathrm{B}}T\ll \Delta E$\ is
described with $X=0$ and $Z=1$, which means replacing the hyperbolic tangent
in Eq.~(\ref{alfagr}) with the unity. The formula~(\ref{alfagr}) for the
ground state obtained by differentiating the probability $P_{\downarrow }$,
Eqs.~(\ref{flux_qb1}-\ref{tan(a)_full}), coincides with the earlier obtained
results (see Eqs.~(3-4) in Ref.~\cite{Ilichev04}). The resulting tank phase
shift is shown in Fig.~\ref{Fig:gr_st_vs_T} for the following parameters
taken from Ref.~\cite{Grajcar04}: $\Delta /h=1.3$\ GHz, $I_{\mathrm{p}}\Phi
_{0}/h=930$\ GHz, $\omega _{0}/2\pi =32.675$\ MHz, $LI_{\mathrm{p}}/\Phi
_{0}=0.0055$, $M/L=0.725$, $Q_{0}=725$, $k=0.02$.

The accurate account of $Z$\ in Eq.~(\ref{alfagr}) allows to describe both
the suppression and widening of the zero-bias dip (that is at $f_{\mathrm{dc}%
}=0$) as it was experimentally demonstrated in Ref.~\cite{Grajcar04}.
Indeed, the suppression of the zero-bias dip (at $f_{\mathrm{dc}}=0$) is
described by the first term in Eq.~(\ref{alfagr}). The widening is due to
the second term that comes from differentiating the hyperbolic tangent; this
term becomes relevant for temperatures larger than $\Delta $, and results in
the exponential rise of the width for $T>T^{\ast }=\Delta /k_{\mathrm{B}}$,
as demonstrated in the inset in Fig.~\ref{Fig:gr_st_vs_T}.

\subsection{Resonant transitions in the charge qubit}

\begin{figure}[t]
\includegraphics[width=8.5cm]{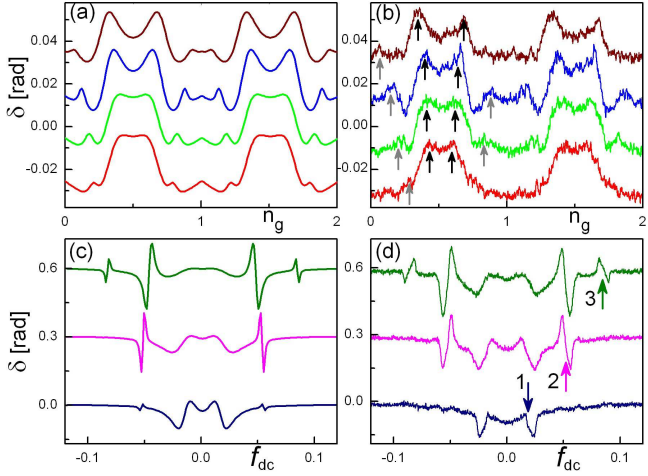}
\caption{(Color online) \textbf{Resonant excitation of the charge qubit
probed by the tank circuit. }The phase shift $\protect\delta $ of the tank
circuit coupled to the charge qubit, calculated theoretically (left) and
measured (right). Panels (a) and (b) show the dependence on the gate
voltage, while in (c) and (d) the dependence on the flux is demonstrated.
Black and gray arrows in (c) demonstrate the positions of 1- and 2-photon
resonant transitions, and the arrows in (d) mark 1-, 2-, and 3-photon
excitations.\ \protect\cite{Shnyrkov06}}
\label{Fig:ShSh}
\end{figure}

In Ref.~\cite{Shnyrkov06} the resonant excitation of the interferometer-type
(split-junction) charge qubit was demonstrated experimentally and described
theoretically. In accordance with the formula (\ref{ch_qb}) one expects the
resonances to appear differently when either first or the second term is
dominated. To demonstrate this, in Fig.~\ref{Fig:ShSh} we plot the
dependence of the tank circuit phase shift $\delta $ both as the function of
the dimensionless bias voltage $n_{\mathrm{g}}=C_{\mathrm{g}}V_{\mathrm{g}%
}/2e$ and of the dimensionless magnetic flux detuning $f_{\mathrm{dc}}$. For
the former case the value $f_{\mathrm{dc}}=0$ was taken, where $I_{0}=0$.
This results in disappearance of the second term in Eq.~(\ref{ch_qb}), and
the resonant excitation of the qubit is visualized with the Lorentzian peaks
in Fig.~\ref{Fig:ShSh}(a,b). When the second term is dominant, the
multiphoton transitions in the qubit result in the peak-and-dip structures
in the dependence of the phase shift $\delta $ on the flux, Fig.~\ref%
{Fig:ShSh}(c,d).

Theoretical fitting of the experimental graphs, as for example shown in Fig.~%
\ref{Fig:ShSh}, allows for defining the qubit's parameters, which is the
\textit{multiphoton spectroscopy}. The parameters found were the following:
the Josephson energies for the two junctions $E_{\mathrm{J}1}/h\simeq 40$
GHz and $E_{\mathrm{J}2}/h\simeq 34.5$ GHz, the island's Coulomb energy $E_{%
\mathrm{C}}/h\simeq 5$ GHz; the relaxation and decoherence rates $\Gamma _{%
\mathrm{relax}}/(E_{\mathrm{C}}/h)=0.03$ and $\Gamma _{\phi }/(E_{\mathrm{C}%
}/h)=0.05$, which correspond to the following relaxation and decoherence
times: $T_{\mathrm{relax}}=\Gamma _{\mathrm{relax}}^{-1}\simeq 7$ ns and $%
T_{\phi }=\Gamma _{\phi }^{-1}\simeq 4$ ns.

Figure~\ref{Fig:ShSh} also demonstrates how the position of the resonances
depend on the driving frequency $\omega $ and how the multiphoton resonances
appear with increasing the driving power $n_{\mathrm{ac}}$. Namely, first,
in Fig.~\ref{Fig:ShSh}(a,b) the varied parameter is the frequency $\omega
/2\pi $, which from the bottom to top curves is $6.5$, $7.1$, $8.1$, and $%
9.1 $ GHz; the driving power is the same for all figures $n_{\mathrm{ac}%
}\simeq 0.3$ and the flux was fixed at $\delta =\pi $. And, second, in Fig.~%
\ref{Fig:ShSh}(c,d) the curves correspond to the varied parameter driving
power: in experiment being power of excitation (from bottom to top: $-80$, $%
-60$, $-57$ dB) and in theory being amplitude $n_{\mathrm{ac}}$ (from bottom
to top: $0.1$, $0.2$, $0.4$); the frequency there was fixed, $\omega /2\pi
=7 $ GHz.

\subsection{One- and multiphoton transitions in the flux qubit}

As we have seen in Section 3, both the tank voltage phase shift $\delta $
and amplitude $v$ can be used to monitor the resonant excitation of a
superconducting qubit. In Fig.~\ref{Fig:ShSh} we demonstrated this with the
observation of the phase shift $\delta $ of the tank circuit coupled to the
charge qubit. Now we consider one- and multi-photon resonant excitations of
a flux qubit, and the nonmonotonic dependence of the tank voltage amplitude $%
v$ will visualize the resonant transitions in the qubit.

\begin{figure}[t]
\includegraphics[width=8cm]{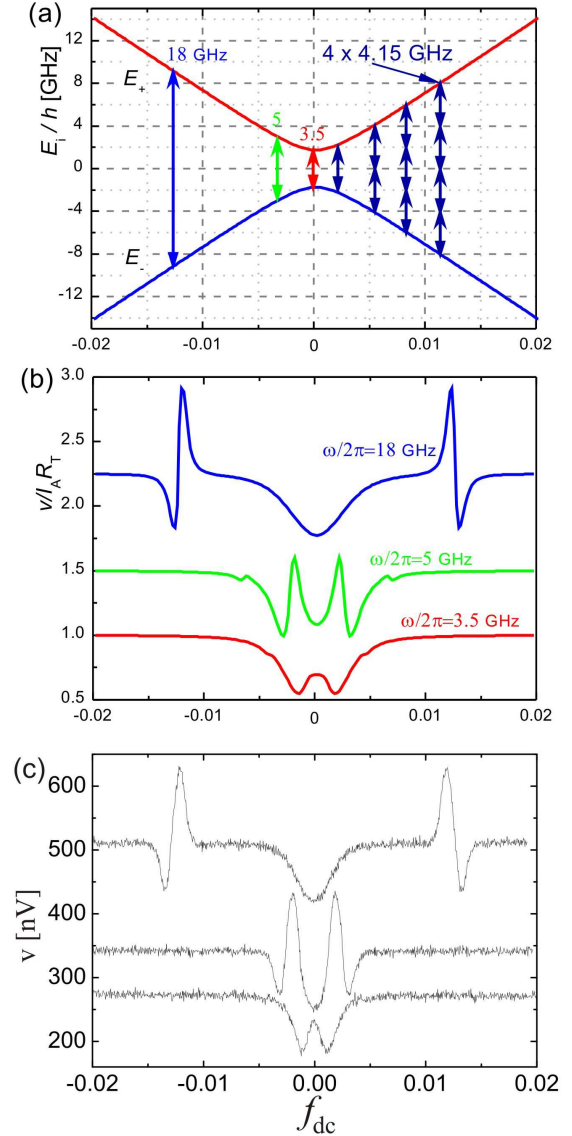}
\caption{(Color online) \textbf{Low-amplitude one-photon resonant excitation
of a flux qubit.} (a) Energy levels $E_{\pm }(f_{\mathrm{dc}})$ matched by
the driving at frequencies shown by the numbers and the arrows of the
respective length. (b) and (c) Theoretically calculated and experimentally
measured amplitude of the tank voltage $v$ versus flux detuning $f_{\mathrm{%
dc}}$ for different driving frequencies. (The upper curves are shifted
vertically.) The one-photon excitations at $\protect\omega /2\protect\pi =18$%
, $5$, and $3.5$ GHz, demonstrated in (b) and (c), are explained by the
arrows to the left in the energy diagram (a), while the arrows to the right
of the length $\protect\omega /2\protect\pi =4.15$ GHz explain the
multiphoton resonances in Fig.~\protect\ref{Fig:multiph_1qb}. \protect\cite%
{ShevchenkoPG08, Izmalkov08} }
\label{Fig:volt_ampl}
\end{figure}

Consider first the \textit{spectroscopical measurement}, where the flux
qubit is driven with the low-amplitude AC flux. We expect resonant
excitation of the qubit when the driving frequency matches the qubit's
energy difference, $\hbar \omega =\Delta E(f_{\mathrm{dc}})$. In the
experimental case the positions of these resonances at a given driving
frequency allow to determine the energy structure of the measured qubit \cite%
{Izmalkov08}.

In Fig.~\ref{Fig:volt_ampl}(b,c) we demonstrate the dependence of the tank
voltage amplitude $v$ on the bias flux $f_{\mathrm{dc}}$ at $\omega _{%
\mathrm{p}}=\omega _{0}$ for different driving frequencies: $\omega /2\pi
=3.5,$ $5,$ and $18$ GHz, which is explained by the energy diagram in \ref%
{Fig:volt_ampl}(a). The results of the related experiment, Ref.~\cite%
{Izmalkov08}, are presented in Fig.~\ref{Fig:volt_ampl}(c). The parameters
for calculations were taken as following: the tunneling amplitude $\Delta
/h=3.5$ GHz, the energy bias $I_{\mathrm{p}}\Phi _{0}/h=700$ GHz, the
temperature $k_{\mathrm{B}}T/h=1.4$ GHz, the relaxation rate $\Gamma
_{1}/h=0.7$ GHz, the dephasing rate $\Gamma _{2}/h=0.7$ GHz, and the value
which describes the coupling between the qubit and the tank circuit $%
k^{2}Q_{0}(LI_{\mathrm{p}}/\Phi _{0})=2.6\cdot 10^{-3}$. The curves were
plotted for the driving amplitudes $f_{\mathrm{ac}}\cdot 10^{3}=1,$ $1.5,$
and $3$ from bottom to top. The phenomenological lagging parameter was taken
$S=0.8$. Figure~\ref{Fig:volt_ampl} demonstrates the effect described in
section 3: for $S\neq 0$ both the phase shift $\delta $ and the amplitude $v$%
\ depend on the qubit's inductance $\mathcal{L}^{-1}$, which results in the
alternation of peak and dip around the location of the resonances.

\begin{figure}[t]
\includegraphics[width=8.5cm]{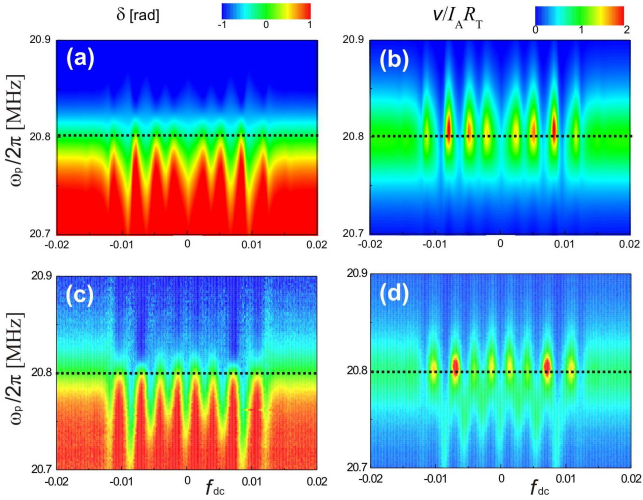}
\caption{(Color online) \textbf{Multiphoton excitations of a flux qubit.}
Theoretically calculated dependence of the phase shift $\protect\delta $ (a)
and the amplitude $v$ (b) on the bias current frequency $\protect\omega _{%
\mathrm{p}}$ and the flux detuning $f_{\mathrm{dc}}$. (c,d) Experimentally
measured phase shift $\protect\delta $ and the amplitude $v$. \protect\cite%
{ShevchenkoPG08}}
\label{Fig:multiph_1qb}
\end{figure}

In Fig.~\ref{Fig:multiph_1qb}(a,b) we present the calculated phase shift $%
\delta $ and the amplitude $v$ as functions of the probe current frequency $%
\omega _{\mathrm{p}}$ and the flux detuning $f_{\mathrm{dc}}$ with the
phenomenological lagging parameter $S$ for the \textit{strongly-driven flux
qubit} with the parameters being the same as for~Fig.~\ref{Fig:volt_ampl}
and with the values for the driving amplitude and frequency: $f_{\mathrm{ac}%
}=8\cdot 10^{-3}$ and $\omega /2\pi =4.15$\ GHz. The top panel presents
theoretical calculations, which is in good agreement with the experimental
observations, presented in the bottom panel, Fig.~\ref{Fig:multiph_1qb}%
(c,d). The dashed white line shows the tank resonance frequency $\omega _{%
\mathrm{p}}/2\pi =\omega _{0}/2\pi =20.8$ MHz. The positions of the
multiphoton resonances is explained by the arrows to the right in the energy
diagram, Fig.~\ref{Fig:volt_ampl}(a), at $\Delta E(f_{\mathrm{dc}})=k\hbar
\omega $ with $k=1$, $2$, $3$, and $4$.


Note that for the lagging parameter close to $1$ (here $S=0.8$) the changes
in the phase shift in Fig.~\ref{Fig:multiph_1qb}(a) are small at the
resonance frequency (along the dotted line at $\omega _{\mathrm{p}}=\omega
_{0}$) while the voltage amplitude in Fig.~\ref{Fig:multiph_1qb}(b) changes
substantially, see formulas (\ref{alpha&v}). And this is actually
demonstrated in Fig.~\ref{Fig:volt_ampl}(b,c). Such changes of the tank
effective resistance or, equivalently, quality factor were studied in Ref.~%
\cite{Grajcar08} for the fully quantum-mechanical model of the
qubit-resonator system. We note that this can be alternatively described
with the semiclassical model, presented here. This model gives results
consistent with the experimental ones, e.g. Figs.~\ref{Fig:volt_ampl} and %
\ref{Fig:multiph_1qb}, which imply the energy transfer between the qubit and
resonator according to Eq.~(\ref{En_trasfer2}). More details about this
energy transfer, known as the \textit{Sisyphus damping and amplification},
can be found in Refs.~\cite{Grajcar08}, \cite{Skinner10}.

\begin{widetext}

\begin{figure}[h]
\includegraphics[width=6cm]{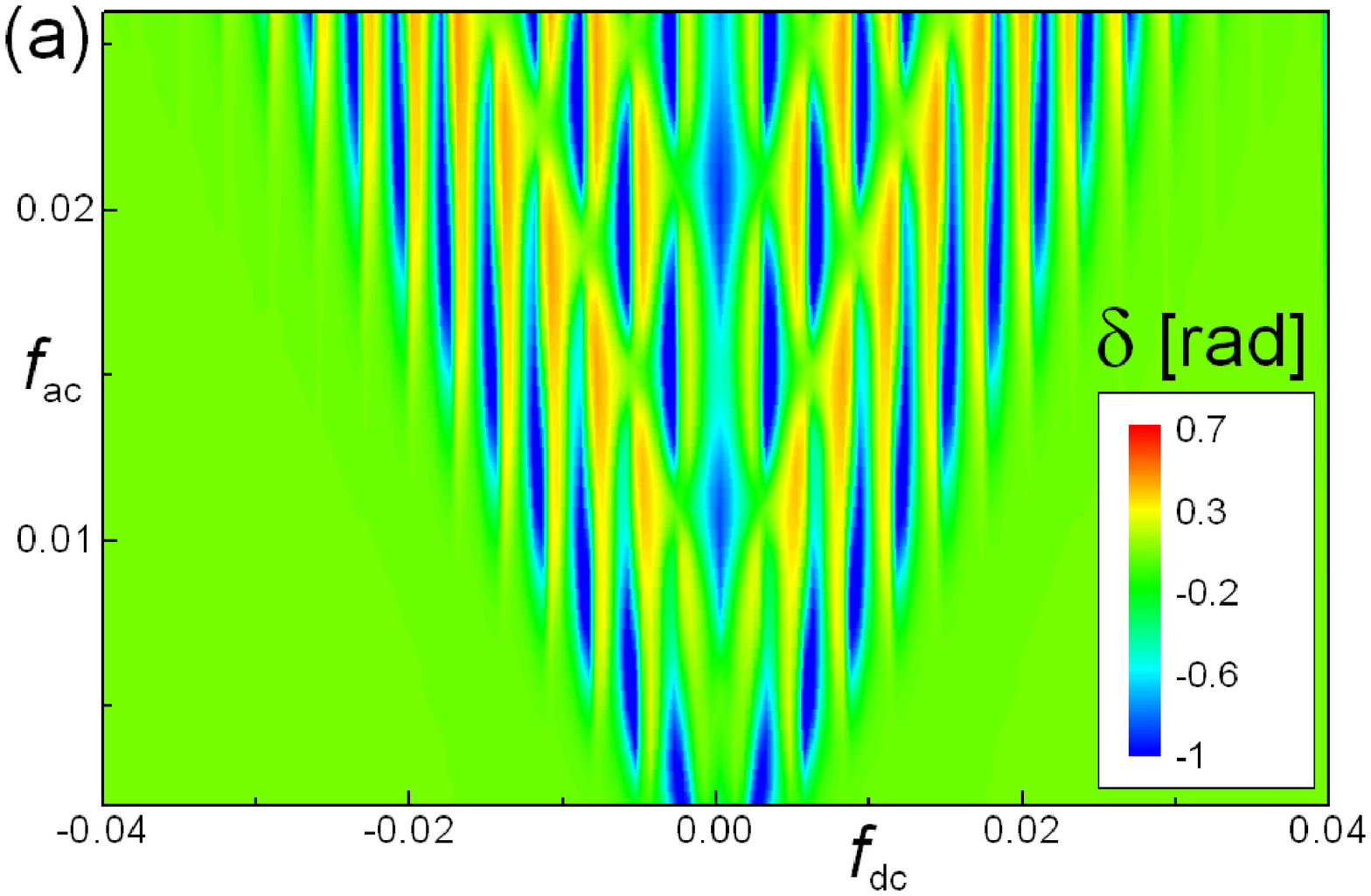} %
\includegraphics[width=6cm]{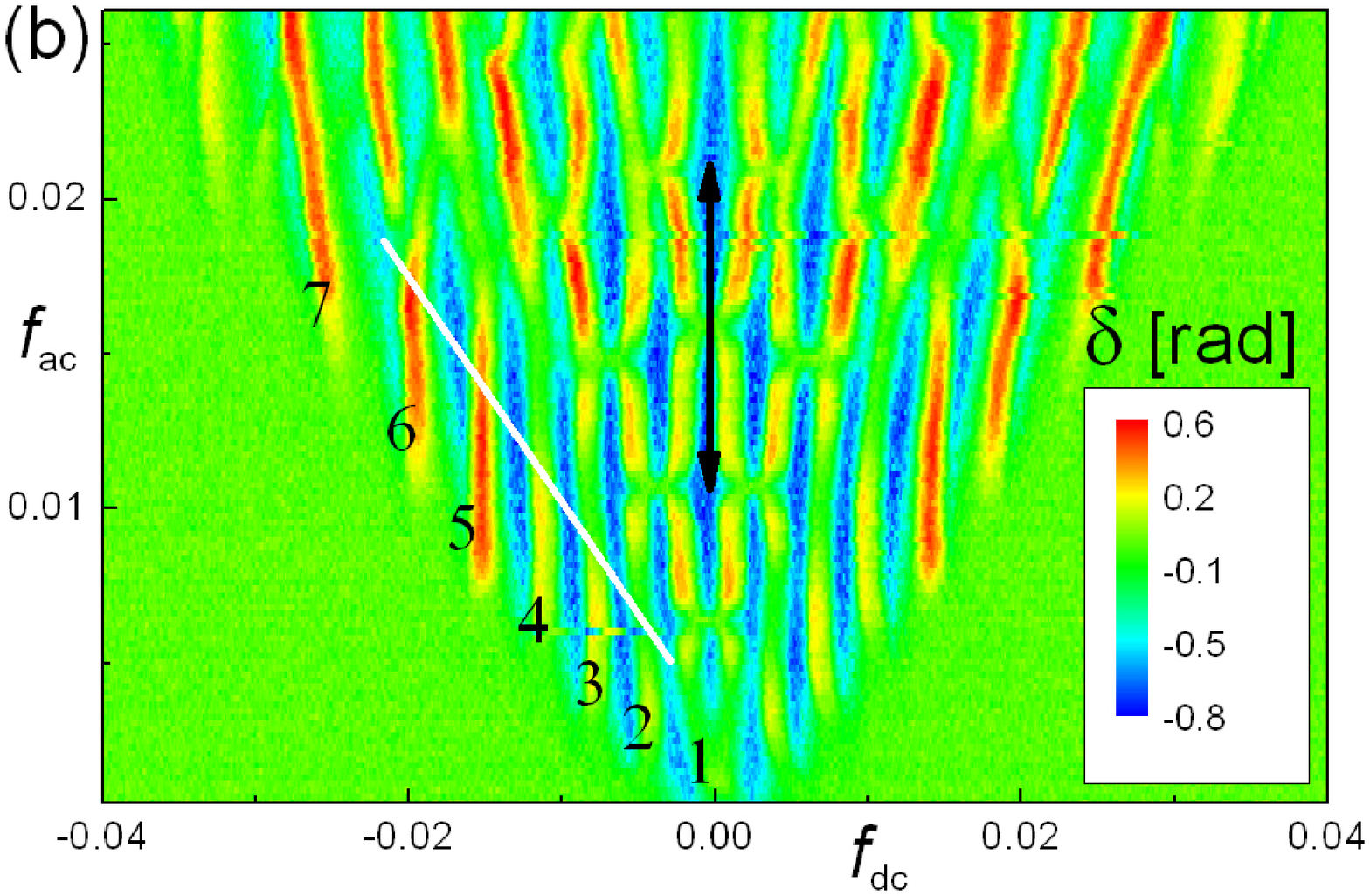}
\caption{(Color online) \textbf{LZS interferometry for the flux
qubit probed by the tank circuit}. The calculated (a) and measured
(b) dependence of the tank phase shift on the flux detuning
$f_{\mathrm{dc}}$ and on the driving flux amplitude
$f_{\mathrm{ac}}$. \protect\cite{ShevchenkoPG08, Izmalkov08}}
\label{Fig:LZSI}
\end{figure}

\end{widetext}

Then, in Fig.~\ref{Fig:LZSI} we present the dependence of the tank voltage
phase shift $\delta $ on the microwave amplitude $f_{\mathrm{ac}}$ and the
DC flux bias $f_{\mathrm{dc}}$. This double quasi-periodical dependence (on
both the energy bias and the driving amplitude) is called the \textit{%
Landau-Zener-St\"{u}ckelberg interferogram} \cite{Shevchenko10}. The
parameters were taken the same as for Fig.~\ref{Fig:volt_ampl} and $\omega
/2\pi =4.15$\ GHz. The left panel in Fig.~\ref{Fig:LZSI} presents the
theoretical interferogram from Ref.~\cite{ShevchenkoPG08} while the right
panel is the experimental one, Ref.~\cite{Izmalkov08}. In Fig.~\ref{Fig:LZSI}
the multiphoton resonances at discrete DC bias $f_{\mathrm{dc}}$\ (which
controls the distance between energy levels) are clearly visible. These
resonances appear when the energy of $k$ photons matches the qubit's energy
levels, $k\hbar \omega \approx \Delta E(f_{\mathrm{dc}})$. The
quasi-periodical character of the dependence on the AC flux amplitude $f_{%
\mathrm{ac}}$ is known as St\"{u}ckelberg oscillations. The comparison of
such graph to the experimental analogue allows the relation of the microwave
power to the AC flux amplitude $f_{\mathrm{ac}}$ to be determined, which is
the calibration of the power. For this, either the estimation of the period
of St\"{u}ckelberg oscillations, shown by the black arrow, or adjusting the
interference pattern slope, shown by the white line, can be used.

\subsection{Interferometry with nanoresonator}

The formalism developed in Sec.~3.3 allows to describe the system of the
driven qubit coupled to the NR. As it was demonstrated in Ref.~\cite%
{Shevchenko11}, two different approaches, called direct and inverse LZS
interferometry, are of interest. In the direct interferometry the qubit
state is probed via the NR's frequency shift, as in Ref.~\cite{LaHaye09},
while in the inverse interferometry the impact of the NR's state on the
qubit's Hamiltonian is studied.

The direct LZS interferometry was calculated in Ref.~\cite{Shevchenko11} as
the resonator's frequency shift $\Delta \omega _{\mathrm{NR}}$ versus the
energy bias $n_{\mathrm{g}}$ and the driving amplitude $n_{\mu }$. The
agreement with the experimental result of Ref. \cite{LaHaye09} demonstrated
that the semiclassical formalism is valid for a description of the
measurable quantities. In Ref.~\cite{Shevchenko11} it is also demonstrated
how the analogous interferogram can be calculated for the qubit-tank circuit
system in relation to the experiment of Ref.~\cite{Sillanpaa06}. Such a
description allows to correctly find the position of the resonance peaks in
the interferogram and to demonstrate the sign-changing behaviour of the
parametric capacitance, which relates to the measurable quantities.

For the formulation of the inverse problem, let us consider the qubit's bias
$\varepsilon _{0}$ as a function of the NR's displacement $x$. For small $%
x\ll \xi $ we have the expansion (\ref{CNR(x)}), which results in the
decomposition of the bias $\varepsilon _{0}(x)\approx \varepsilon _{0}^{\ast
}(n_{\mathrm{g}})+\delta \varepsilon _{0}(x)$, where $\varepsilon _{0}^{\ast
}(n_{\mathrm{g}})=8E_{\mathrm{C}}\left( n_{\mathrm{g}}-1/2\right) $ and $%
\delta \varepsilon _{0}(x)=8E_{\mathrm{C}}n_{\mathrm{NR}}x/\xi $. The
Hamiltonian of the qubit (\ref{H1qb}) with the parameter-dependent bias $%
\varepsilon _{0}(x)$ allows to consider the following problem. Let us assume
that the qubit's state (its wave function, upper level occupation
probability, Rabi frequency, etc.) is known (i.e. this is measured by a
device, which we do not consider here for simplicity). Given the known
qubit's state, the aim is to find the Hamiltonian's parameters. Particularly
interesting is the parameter-dependent bias $\varepsilon _{0}(x)$, which can
give the information about the position and amplitude of the oscillations of
the NR.

And now, in the general context, the \textquotedblleft reverse
engineering\textquotedblright\ problem in the spirit of Refs.~[%
\onlinecite{Garanin02, Berry09}] can be studied, where one is interested in
finding the driving Hamiltonian for a given (desired) final state. On the
other hand, in Ref.~\cite{Shevchenko11} the authors provide the basis for
measuring the NR's position $x$ by means of probing the qubit's state, while
$x=x(t)$ is considered a slow time-dependent function. There, the emphasis
was made on finding optimal driving and controlled offset ($\varepsilon
_{0}^{\ast }$) parameters for the resolution of the small bias component $%
\delta \varepsilon _{0}$. It was assumed that the dynamics of the parameter $%
x$ is slow enough not to be considered during either certain period of the
qubit's evolution or even during the setting the stationary qubit's state.
The aim was to find a sensitive probe for small $\delta \varepsilon _{0}$.
As the ultimate sensitivity, the essential changes of the qubit's state for
small changes of $\delta \varepsilon _{0}$ were required. The problem,
formulated in this way, was solved in Ref.~\cite{Shevchenko11} for different
illustrative driving regimes: one-, double-, and multiple-passage regimes.

\section{5. Multi-qubit systems}

\subsection{Equations for a system of coupled qubits}

The effective Hamiltonian of the system of $n$ coupled flux qubits is
\begin{equation}
H=\sum\limits_{i=1}\left( -\frac{\Delta _{i}}{2}\sigma _{x}^{(i)}-\frac{%
\varepsilon _{i}(t)}{2}\sigma _{z}^{(i)}\right) +\sum\limits_{i,j}\frac{%
J_{ij}}{2}\sigma _{z}^{(i)}\sigma _{z}^{(j)},  \label{H2qbs}
\end{equation}%
where $J_{ij}$ is the coupling energy between qubits, and $\sigma _{x}^{(i)}$%
, $\sigma _{z}^{(i)}$ are the Pauli matrices in the basis $\{\left\vert
\downarrow \right\rangle ,\left\vert \uparrow \right\rangle \}$ of the
current operator in the $i$-th qubit. The current operator is given by: $%
I_{i}=-I_{\mathrm{p}}^{(i)}\sigma _{z}^{(i)},$ with $I_{\mathrm{p}}^{(i)}$
the absolute value of the persistent current in the $i$-th qubit; then the
eigenstates of $\sigma _{z}$\ correspond to the clockwise ($\sigma
_{z}\left\vert \downarrow \right\rangle =-\left\vert \downarrow
\right\rangle $) and counterclockwise ($\sigma _{z}\left\vert \uparrow
\right\rangle =\left\vert \uparrow \right\rangle $) current in the $i$-th
qubit. The tunneling amplitudes $\Delta _{i}$ are assumed to be constants.
The biases $\varepsilon _{i}=2I_{\mathrm{p}}^{(i)}\Phi _{0}f^{(i)}(t)$ are
controlled by the dimensionless magnetic fluxes $f^{(i)}(t)=\Phi _{i}/\Phi
_{0}-1/2$ through $i$-th qubit. These fluxes consist of three components,
\begin{equation}
f^{(i)}(t)=f_{i}+\frac{M_{i}I_{\mathrm{L}}}{\Phi _{0}}+f_{\mathrm{ac}}\sin
\omega t.  \label{fx}
\end{equation}%
Here $f_{i}$ is the adiabatically changing magnetic flux, experimentally
applied by the coil and additional DC lines. The second term describes the
flux induced by the current $I_{\mathrm{L}}$ in the tank coil, to which the $%
i$-th qubit is coupled with the mutual inductance $M_{i}$. And $f_{\mathrm{ac%
}}\sin \omega t$ is the harmonic time-dependent component driving the qubit,
typically applied by an on-chip microwave antenna. Equation (\ref{H2qbs})
can be reduced to the two-qubit system. This system is shown in Fig.~\ref%
{Fig:scheme_2qbs}.

\begin{figure}[t]
\includegraphics[width=7cm]{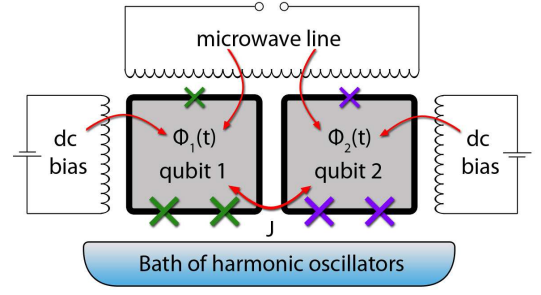}
\caption{(Color online) \textbf{Scheme of two coupled qubits.} The two flux
qubits are coupled to each other, to the dc and $\protect\mu $w lines, as
well as to an unavoidable dissipative environment. The convenient model for
description of the environment is the bath of harmonic oscillators. The
system of two coupled qubits is also assumed to be coupled to the measuring
resonant circuit (which is not shown here), as in Fig.~\protect\ref%
{Fig:qb+LCR}. \protect\cite{Temchenko11}}
\label{Fig:scheme_2qbs}
\end{figure}
To describe the two-qubit system, it is convenient to present the density
matrix in the following form
\begin{eqnarray}
\rho &=&\frac{R_{\alpha \beta }}{4}\sigma _{\alpha }\otimes \sigma _{\beta }=%
\frac{R_{00}}{4}\sigma _{0}\otimes \sigma _{0}+  \notag \\
+\frac{R_{a0}}{4}\sigma _{a}\otimes \sigma _{0} &+&\frac{R_{0b}}{4}\sigma
_{0}\otimes \sigma _{b}+\frac{R_{ab}}{4}\sigma _{a}\otimes \sigma _{b},
\label{dens_mat}
\end{eqnarray}%
which was shown to be suitable for both the definition and the calculation
of the entanglement and other characteristics in multi-qubit system, e.g.
\cite{Schlienz95, Ivanchenko07}. Here $\alpha ,\beta =0,x,y,z$ and $%
a,b=x,y,z $; the summation over twice repeating indices is assumed. The two
vectors $R_{a0}$ and $R_{0b}$, so-called coherence vectors or Bloch vectors,
determine the properties of the individual qubits, while the tensor $R_{ab}$
(the correlation tensor) accounts for the correlations.

The important characteristic of the state of the coupled-qubits system is
its entanglement. There are different approaches to the quantification of
the entanglement \cite{Love07}. One of the often used possibilities is the
so-called concurrence \cite{Li08}. Another convenient for calculations
approach is to introduce the measure of entanglement as following \cite%
{Schlienz95}
\begin{equation}
\mathcal{E}=\frac{1}{3}\mathrm{Tr}\left( M^{T}M\right) ,\text{ }%
M_{ab}=R_{ab}-R_{a0}R_{0b}.  \label{E}
\end{equation}%
This entanglement measure fulfills certain requirements, in particular, $%
\mathcal{E}=0$ for any product state and $\mathcal{E}=1$ for any pure state
with vanishing Bloch vectors $R_{a0}$ and $R_{0b}$, corresponding to maximum
entangled states.

To describe dynamics of the density matrix we will first disregard the
relaxation processes. This can be described by the Liouville equation, $%
i\hbar \dot{\rho}=\left[ H,\rho \right] $, which is generally speaking a
complex equation. To deal with the Liouville equation, it is convenient to
use the parametrization of the density matrix as described by Eq.~(\ref%
{dens_mat}). Due to the hermiticity and normalization of the density matrix,
$R_{\alpha \beta }$ are real numbers and $R_{00}=1$. Then the Liouville
equation can be written in the form of the system of $15$ equations for $%
R_{\alpha \beta }$ \cite{Temchenko11}
\begin{eqnarray}
\dot{R}_{i0} &=&\epsilon _{mni}B_{m}^{(1)}R_{n0}+\epsilon _{3ni}JR_{n3},
\notag \\
\dot{R}_{0j} &=&\epsilon _{mnj}B_{m}^{(2)}R_{0n}+\epsilon _{3nj}JR_{3n},
\notag \\
\dot{R}_{ij} &=&\epsilon _{mni}B_{m}^{(1)}R_{nj}+\epsilon
_{mnj}B_{m}^{(2)}R_{in}  \notag \\
&+&\delta _{j3}\epsilon _{3ni}JR_{n0}+\delta _{i3}\epsilon _{3nj}JR_{0n},
\label{sys_15_eqs}
\end{eqnarray}%
where $\mathbf{B}^{(i)}=\left( -\Delta _{i},0,-\varepsilon _{i}\right) $ and
$\epsilon _{mni}$ is the Levi-Civita symbol.

Consider now the measurable value, which is the resonator's phase shift. As
we discussed in Sec. 2, it relates to the effective inductance of qubits
system. The formula obtained for single qubits can be generalized for the
two-qubit system \cite{Grajcar05, Shevchenko08}. Then for the case of
low-quality~qubits, when their characteristic times are smaller than the
tank's period, at the resonance frequency ($\xi _{0}=0$), expression for the
phase shift $\delta $ in terms of the parametric inductances $L_{\mathrm{q}%
}^{(i)}$ can be written as following%
\begin{eqnarray}
\tan \delta  &\approx &Q_{0}\sum_{i=1,2}\frac{L_{0}}{L_{\mathrm{q}}^{(i)}},%
\text{ }  \label{alpha_4_2} \\
\frac{L_{0}}{L_{\mathrm{q}}^{(i)}} &=&k^{2}\frac{L_{i}}{\mathcal{L}_{i}}%
\text{ },\text{ }\mathcal{L}_{i}^{-1}=\left( \frac{\partial }{\partial \Phi
_{a}}+\frac{\partial }{\partial \Phi _{b}}\right) I_{\mathrm{qb}}^{(i)}.
\notag
\end{eqnarray}%
In what follows this expression will be used to calculate the phase shift $%
\delta $, which maps the qubits' state.

\subsection{Weak-driving spectroscopy}

In Sec.~4 we have considered how the measurements of the single qubits allow
to determine their parameters: the tunneling amplitudes $\Delta \ $and the
persistent currents $I_{\mathrm{p}}$. It was demonstrated~\cite{Izmalkov08}
that for defining the parameters of single and multiple-qubits systems both
the ground-state measurements and excited-state spectroscopy can be used;
the consistency of the results of the two approaches was shown. Now we will
demonstrate this for the case of the system of two coupled flux qubits
described by the Hamiltonian (\ref{H2qbs}).

First, the one-qubit parameters are defined. For this, suppose qubit $a$ is
the one biased far from its degeneracy point in such a way that $\varepsilon
_{a}$ is large in comparison with the other energy variables. Then, qubit $a$
has a well defined ground state with averaged spin variables $\left\langle
\sigma _{z}^{(a)}\right\rangle =1$ and $\left\langle \sigma
_{x}^{(a)}\right\rangle =0$ which can be averaged out of the two-qubit
Hamiltonian (\ref{H2qbs}) reducing it to: $H_{\mathrm{2qbs,red}}=-\Delta
_{b}\sigma _{x}^{(b)}/2-(\varepsilon _{b}-J)\sigma _{z}^{(b)}/2$. Apart from
the offset in the bias term due to the coupling, this is identical to the
single qubit Hamiltonian. This offset can be easily compensated and measured
allowing the determination of the coupling energy $J$ \cite{Grajcar05}. The
qubit parameters, $\Delta _{b}$ and $I_{\mathrm{p}}^{(b)}$, are determined
from either the ground-state measurement or the excited-state spectroscopy,
as it is described in Sec.~4. Analogously, biasing qubit $b$ far from the
degeneracy point the parameters for qubit $a$, $\Delta _{a}$ and $I_{\mathrm{%
p}}^{(a)}$, can be determined. Next, the coupling energy $J$ was determined
from the offset of the qubit dips from the $\Phi _{a/b}=0$ lines, visible in
the pure ground-state measurements presented in Fig.~\ref%
{Fig:spectroscopy_2qbs}(a).

\begin{figure}[t]
\includegraphics[width=8cm]{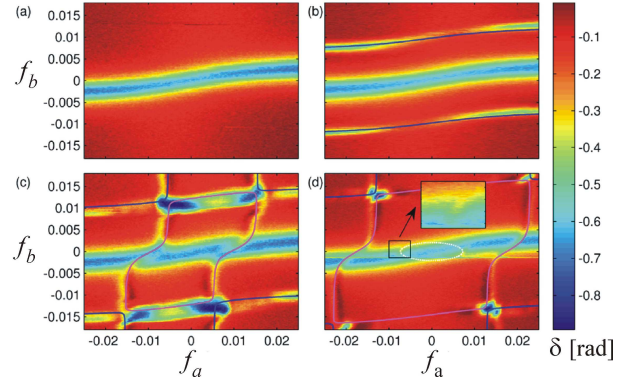}
\caption{(Color online) \textbf{Spectroscopy of the two-qubit system.} The
measured dependence of the phase shift $\protect\delta $ on the flux biases $%
f_{a}$ and $f_{b}$: (a) ground-state measurement (without microwave
excitation). (b,c,d) with weak microwave excitation at the driving
frequencies $\protect\omega /2\protect\pi =$~$14.1$, $17.6$ and $20.7$ GHz.
\protect\cite{Izmalkov08}}
\label{Fig:spectroscopy_2qbs}
\end{figure}

Then the qubits were driven by magnetic fluxes $\Phi _{\mathrm{ac}}\sin
\omega t$ with weak driving amplitudes and various driving frequencies.
There, we expect the position of the resonant transitions from energy level $%
E_{j}$ to an overlying level $E_{i}$ determined by the one-photon relation: $%
\Delta E_{ij}(\Phi _{a},\Phi _{b})\approx \hbar \omega $, which appears when
the distance between the energy levels $\Delta E_{ij}=E_{i}-E_{j}$ is
matched by the photon energy $\hbar \omega $. In Fig.~\ref%
{Fig:spectroscopy_2qbs}(b) a frequency in-between both qubit gaps ($\Delta
_{b}<\hbar \omega <\Delta _{a}$) was used and therefore only the transitions
to the first excited state are visible. For higher frequencies, also the
second and third excited states become visible as can be seen in subfigures
(c) and (d). The theoretically calculated contour lines are superposed in
Fig.~~\ref{Fig:spectroscopy_2qbs}(b-d) for three different frequencies for
which the condition, $\Delta E_{ij}(\Phi _{a},\Phi _{b})=\hbar \omega $, is
fulfilled; the energy levels $E_{i}=E_{i}(\Phi _{a},\Phi _{b})$ were found
by diagonilizing the Hamiltonian. From the fitting procedure the following
parameters were found: the tunneling amplitudes $\Delta _{a(b)}/h=15.8(3.5)$
GHz, the energy biases $I_{\mathrm{p}}^{a(b)}\Phi _{0}/h=375(700)$ GHz [$I_{%
\mathrm{p}}^{a(b)}=120(225)$ nA], the inter-qubit coupling $J/h=3.8$ GHz,
and the value which describes the coupling between the qubits and the tank
circuit $\Xi _{a(b)}=1.4(2.6)\cdot 10^{-3}$, where $\Xi
_{i}=k_{i}^{2}Q_{0}(L_{i}I_{\mathrm{p}}^{(i)}/\Phi _{0})$.

\subsection{Direct and ladder-type multiphoton transitions}

We now consider the multiphoton excitations of a system of two strongly
driven coupled flux qubits. We will describe the effects of resonant
excitation in the system in terms of its energy structure, entanglement
measure, and the observable tank circuit phase shift. Then we will present
results for the multiphoton excitation of two types: direct (when multiple
photon energy $k\hbar \omega $ matches the energy level difference $\Delta
E_{ij}$) and ladder-type (when the transition happens via an intermediate
level). We will demonstrate how this can be used for creating the inverse
population in the dissipative two-qubit system.

\begin{figure}[t]
\includegraphics[width=8.5cm]{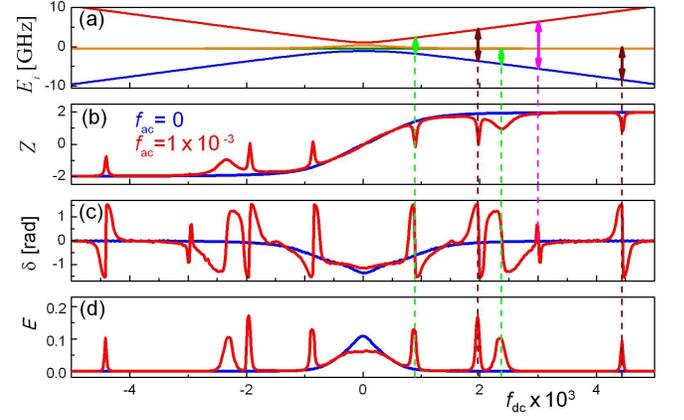}
\caption{(Color online) \textbf{Characterizing strongly-driven two-qubit
system}$.$ Calculated and plotted as functions of the bias $f_{\mathrm{dc}}$
are four energy levels (a), total probability of the currents in two qubits
to flow clockwise $Z$ (b), the tank circuit voltage phase shift $\protect%
\delta $ (c), the entanglement measure $\mathcal{E}$ (d). \protect\cite%
{ShevchenkoPG08}}
\label{Fig:multiph_2qbs}
\end{figure}

To describe the system of two qubits subjected to the strong driving, the
following values were calculated: the energy levels (by diagonalizing the
stationary Hamiltonian), the density matrix $\rho $ (by solving the
Liouville equation), the observable tank circuit phase shift $\delta $
(which is defined with the effective inductance of the qubits), and the
entanglement measure $\mathcal{E}$\ by making use of Eqs.~(\ref{E}-\ref%
{alpha_4_2}). In this way graphs in Fig.~\ref{Fig:multiph_2qbs} were
calculated for the set of parameters of the two-qubit system realized in
Ref.~\cite{Izmalkov04a}: $\Delta _{a}/h=1.2$ GHz, $\Delta _{b}/h=0.9$ GHz, $%
I_{\mathrm{p}}^{(a,b)}\Phi _{0}/h=$ $990$ GHz, $J/h=0.84$ GHz, $\Xi
_{a,b}=1.8\cdot 10^{-3}$, and the driving frequency was taken $\omega /2\pi
=4$\ GHz; also the change of the DC flux here was assumed symmetrical: $%
f_{a}=f_{b}\equiv f_{\mathrm{dc}}$. For simplicity here the relaxation
processes were ignored (and we will pay special attention to this below) and
we consider the case when the characteristic measurement time $T_{\mathrm{p}%
}=2\pi /\omega _{\mathrm{p}}$ is larger than the characteristic times of the
dynamics of the qubit. Then the tank circuit actually probes the incoherent
mixture of qubit's states and the time-averaged values of phase shift and
entanglement should be considered.

When the energy of $k$ photons ($k\hbar \omega $) matches the energy
difference between any two levels $E_{j}$ and $E_{i}$, the resonant
excitation to the upper level is expected. Respectively, the arrows of the
length $4$, $8$, and $12$\ GHz show the places of possible one-, two-, and
three-photon excitations. The time-averaged total probability of the
currents in two qubits to flow clockwise, $Z=R_{03}+R_{30}$, is shown in
Fig.~\ref{Fig:multiph_2qbs}(b) to experience resonant excitation. The
resonances appear as peak-and-dip structures in the phase shift dependence
in Fig.~\ref{Fig:multiph_2qbs}(c). The time-averaged entanglement measure $%
\mathcal{E}$ in a resonance increases due to the formation of the
superposition of states, Fig.~\ref{Fig:multiph_2qbs}(d); this provides a
method to control and probe the entanglement.


\begin{figure}[t]
\includegraphics[width=8.5 cm]{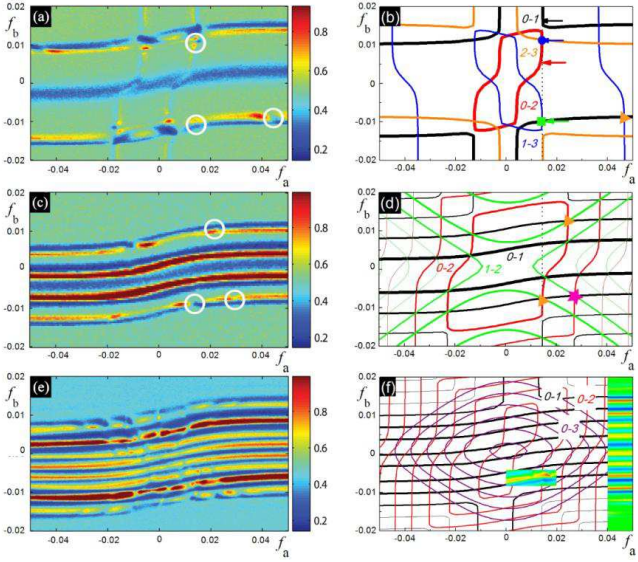}
\caption{(Color online) \textbf{Imaging the} \textbf{multiphoton transitions
in the two-qubit system.} The resonant excitation of the qubits system is
visualized by the tank voltage amplitude, (a, c, e). The position of the
resonant transitions can be understood by comparing with the respective
energy contour lines (b, d, f). \protect\cite{Ilichev10}}
\label{Fig:multiphoton-multilevel}
\end{figure}

The experimental study of the strongly driven system of two coupled flux
qubits is presented in Fig.~\ref{Fig:multiphoton-multilevel}. The left panel
is the measured voltage amplitude of the tank as a function of qubit biases $%
f_{\mathrm{a}}$ and $f_{\mathrm{b}}$. The driving frequencies from top to
bottom were $\omega /2\pi =17.6$, $7.0$, and $4.1$ GHz. The multiphoton
resonances at $\Delta E_{ij}(f_{a},f_{b})\approx k\hbar \omega $\ are
visualized with the ridge-trough lines. We note that the resonance
ridge-trough lines are disturbed with increasing or decreasing the signal;
some of these changes are shown with white circles. This means changing the
effective Josephson inductance in these points. The experimental results can
be clearly understood by comparing them with the energy contour lines,
calculated by diagonalizing Hamiltonian (\ref{H2qbs}) and presented in the
right panel of the figure. There, numbers $k-j$ next to the lines mean that
the line relates to the energy difference $E_{j}-E_{k}$.

\begin{figure}[t]
\includegraphics[width=6cm]{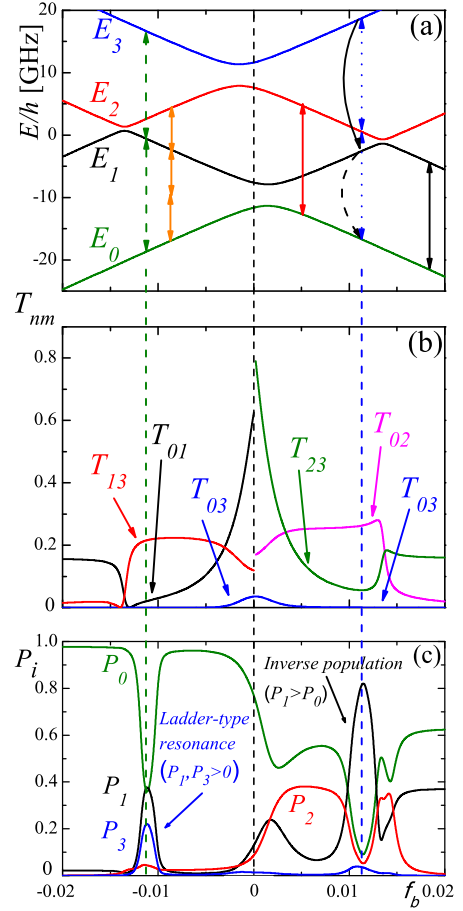}
\caption{(Color online) \textbf{Ladder-type transitions in the
two-qubit system. }Calculated as functions of the flux $f_{\mathrm{b}}$ (at $f_{%
\mathrm{a}}=0.015$): the energy levels (a), transition matrix elements $%
T_{nm}$ (b), the occupation probabilities $P_{i}$ (c). \protect\cite%
{Ilichev10}}
\label{Fig:levels_and_populations_2qbs}%
\end{figure}

Consider now these multiphoton features in more details. In Fig.~\ref%
{Fig:multiphoton-multilevel}(b) the black and red lines show the positions
of the expected resonant excitations from the ground state to the first and
to the second excited states respectively; the blue and orange lines are the
contour lines for the possible excitations from the first and from the
second excited state to the third excited state. In Fig.~\ref%
{Fig:levels_and_populations_2qbs}(a) the energy levels are plotted at the
fixed value of the bias flux through qubit $a$, $f_{\mathrm{a}}$, as a
function of the bias flux through qubit $b$, $f_{\mathrm{b}}$. The arrows
are introduced to match the energy levels with the driving frequencies $%
\omega /2\pi =17.6$ GHz and $7.0$ GHz . The black and red arrows in both
Fig.~\ref{Fig:levels_and_populations_2qbs}(a) and Fig.~\ref%
{Fig:multiphoton-multilevel}(b) show the position of one-photon transitions
to the first and the second excited levels. The double green and blue arrows
in Fig.~\ref{Fig:multiphoton-multilevel} show the position of the two-photon
processes, where the excitation by the first photon creates the population
of the first and the second levels and the second photon excites the system
to the upper level. These two-photon excitations happen via intermediate
levels; compare the position of these expected resonances in Fig.~\ref%
{Fig:multiphoton-multilevel}(b) shown with the blue circle and green square.
The orange triangle in Fig.~\ref{Fig:multiphoton-multilevel}(b) points the
ladder-type three-photon excitation, with one photon to the first excited
level and then with two photons to the upper level.


Analogous considerations allow to see in Fig.~\ref%
{Fig:multiphoton-multilevel}(c) and (d) one- and two-photon resonant
excitations to the first excited level for the driving frequency $\omega
/2\pi =7$ GHz. The two-photon resonant excitation is direct and happen
without any intermediate level. The higher level excitations via the first
excited state appear due to three- and four-photon excitations, as shown
with orange triangles and pink asterisk. In Fig.~\ref%
{Fig:multiphoton-multilevel}(e) the response of the two-qubit system at $%
\omega /2\pi =4.1$ GHz exhibits 1- to 4-photon excitations to the first
excited state, which can be recognized by comparing with the black lines in
Fig.~\ref{Fig:multiphoton-multilevel}(f). Numerous upper level excitations
via the first excited level appear as the changes of the signal along these
lines.

The transition rates can be quantified by the absolute value of the matrix
element of the perturbation between the states $\left\vert
E_{m}\right\rangle $ and $\left\vert E_{n}\right\rangle $%
\begin{equation}
T_{nm}=\left\vert \left\langle E_{n}\right\vert \widehat{v}\left\vert
E_{m}\right\rangle \right\vert ^{2},\text{ \ }\widehat{v}=\frac{1}{I_{%
\mathrm{p}}^{(\mathrm{b})}}\left( I_{\mathrm{p}}^{(\mathrm{a})}\widehat{%
\sigma }_{z}^{(\mathrm{a})}+I_{\mathrm{p}}^{(\mathrm{b})}\widehat{\sigma }%
_{z}^{(\mathrm{b})}\right) ,
\end{equation}%
divided by the factor $I_{\mathrm{p}}^{(\mathrm{b})}\Phi _{0}f_{\mathrm{ac}}$%
. The transition matrix elements in Fig.~\ref%
{Fig:levels_and_populations_2qbs}(b) explain the ladder-type excitations in
Fig.~\ref{Fig:multiphoton-multilevel}(b). Two points, marked by the vertical
dashed green and blue lines in Fig.~\ref{Fig:levels_and_populations_2qbs}
describe respectively two interesting situations. To the right (see along
the blue line) the transition element between the higher two levels ($E_{2}$
and $E_{3}$) is smaller than between the lower two levels ($E_{0}$ and $E_{2}
$), $T_{02}\gg T_{23}\gg T_{03}$. In contrast, to the left (see along the
green line) the transition element between the higher two levels ($E_{1}$
and $E_{3}$) is larger than between the lower two levels ($E_{0}$ and $E_{1}$%
), $T_{13}\gg T_{01}\gg T_{03}$. In both cases the probability of the direct
excitation to the highest level is very small, which means that the
transitions are induced due to the ladder-type mechanism.

The ladder-type transitions and the population inversion can be also
illustrated by calculating the energy level occupation probabilities by
solving the Bloch-Redfield equation (see the next subsection for more
details); figure~\ref{Fig:levels_and_populations_2qbs}(c) was calculated
with the driving frequency $\omega /2\pi =17.6$ GHz and amplitude $f_{%
\mathrm{ac}}=4\times 10^{-3}$. First, the ladder-type resonant excitation
takes place to the left, where the upper level occupation probability $P_{3}$
is of the same order as the intermediate level occupation probability $P_{1}$%
. Second, the inverse population\textit{\ }appears to the right, where the
upper level occupation probability $P_{1}$ is larger than the ground state
probability $P_{0}$, see also Refs.~\cite{Astafiev07, You07, Berns08, Sun09}
for the study of the population inversion in the systems with single
Josephson-junction qubits. These two phenomena are similar to those which
exhibit atoms in the laser field~\cite{Vitanov01}. Furthermore, the
expectation value of the current in $i$-th qubit is calculated with the
reduced density matrix: $I_{\mathrm{qb}}^{(i)}=-I_{\mathrm{p}}^{(i)}Sp(\rho
\sigma _{z}^{(i)})$. The results of the calculations are also presented as
the color insets in Fig.~\ref{Fig:multiphoton-multilevel}(f) for the
following parameters: the strength of dissipation $\alpha =0.1$ and the
driving amplitude $f_{\mathrm{ac}}=8\times 10^{-3}$.

\subsection{Lasing in the two-qubit system}

Consider now the influence of the dissipation on the dynamics of a two-qubit
system. For this the Bloch-Redfield formalism will be used. The strong
dependence of the inter-level relaxation rates on the controlling magnetic
fluxes will be demonstrated for the realistic system. This allows to propose
several mechanisms for lasing in this four-level system~\cite{Temchenko11}.

\begin{figure}[t]
\includegraphics[width=6.2cm]{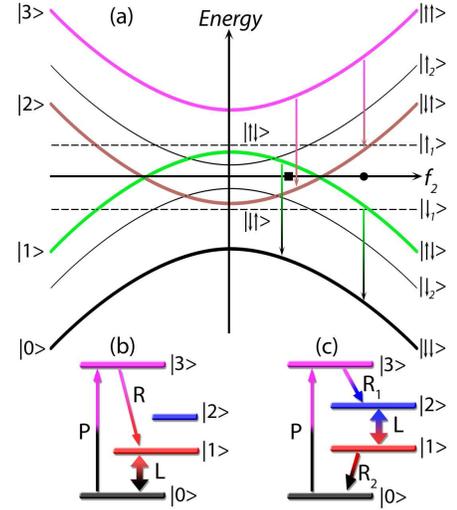}
\caption{(Color online) \textbf{Energy level structure with }$J=0$. (a)
One-qubit and two-qubit energy levels as functions of the magnetic flux $%
f_{b}$ at fixed flux $f_{a}$. The arrows show the fastest relaxation, which
is assumed to relate to the qubit $a$. (b) and (c) Schemes for three- and
four-level lasing at $f_{b}=f_{b\mathrm{L}}$ and $f_{b}=f_{b\mathrm{R}}$.
The driving magnetic flux pumps (P) the upper level; fast relaxation (R)
creates the population inversion; the two operating levels can be used for
lasing (L). \protect\cite{Temchenko11}}
\label{Fig:levels_2qbs}
\end{figure}

For identification of the level structure and understanding different
transition rates it is instructive to start from considering the case of two
non-interacting qubits, that is $J=0$. In this simplified situation, the
energy levels of the system of two qubits consist of the pair-wise summation
of single-qubit levels,
\begin{equation}
E_{i}^{\pm }=\pm \frac{\Delta E_{i}}{2}=\pm \frac{1}{2}\sqrt{\varepsilon
_{i}^{(0)2}+\Delta _{i}^{2}}.  \label{DE}
\end{equation}%
In Fig.~\ref{Fig:levels_2qbs}(a) the energy levels are plotted as a function
of the partial bias in the second qubit $f_{b}$, fixing the bias in the
first qubit $f_{a}$. Then the single-qubit energy levels appear as the
horizontal lines at $E_{a}^{\pm }=\pm \frac{1}{2}\sqrt{\varepsilon
_{a}^{(0)2}+\Delta _{a}^{2}}$ and as the parabolas at $E_{b}^{\pm
}(f_{b})=\pm \frac{1}{2}\sqrt{\varepsilon _{b}^{(0)}(f_{b})^{2}+\Delta
_{b}^{2}}$. For the lasing the hierarchy of the relaxation times is
required. For this it is natural to assume that the relaxation in the first
qubit is much faster than in the second qubit. This allows to consider
three- and four-level lasing schemes in Fig.~\ref{Fig:levels_2qbs}(b,c).

As a next step, the interaction of the qubits, $J\neq 0$, should be
considered. To describe the relaxation in this system, the operators are
converted to the basis of eigenstates of the unperturbed Hamiltonian. In
this representation $H_{0}^{\prime }=S^{-1}H_{0}S$ is the diagonal matrix;
the unitary matrix $S$ consists of eigenvectors of the unperturbed
Hamiltonian; the excitation operator $V(t)$ is converted as following
\begin{equation}
V^{\prime }(t)=S^{-1}V(t)S=\sum_{i=1,2}-\frac{1}{2}\widetilde{\varepsilon }%
_{i}(t)\tau _{z}^{(i)},\text{ }\tau _{z}^{(i)}=S^{-1}\sigma _{z}^{(i)}S.
\label{V'}
\end{equation}

The dissipative environment can be described as the thermostat, for which
the convenient model is the bath of harmonic oscillators, see Fig.~\ref%
{Fig:scheme_2qbs}. Within the Bloch-Redfield formalism, the Liouville
equation for the quantum system interacting with the bath is transformed
into the master equation for the reduced system's density matrix $\rho (t)$.
Then the master equation for the density matrix of our driven system can be
written in the energy representation as following \cite{Blum, Weiss}
\begin{equation}
\dot{\rho}_{ij}=-i\omega _{ij}\rho _{ij}-\frac{i}{\hbar }\left[ V^{\prime
},\rho \right] _{ij}+\delta _{ij}\sum_{n\neq j}\rho _{nn}W_{jn}-\gamma
_{ij}\rho _{ij},  \label{M_eqn}
\end{equation}%
where $\omega _{ij}=(E_{i}-E_{j})/\hbar $, and the relaxation rates $W_{mn}=2
$Re$\Gamma _{nmmn}$ and
\begin{equation}
\gamma _{mn}=\sum_{r}\left( \Gamma _{mrrm}+\Gamma _{nrrn}^{\ast }\right)
-\Gamma _{nnmm}-\Gamma _{mmnn}^{\ast }  \label{gmn}
\end{equation}%
are defined by the relaxation tensor $\Gamma _{lmnk}$, which is given by the
Fermi Golden rule. As it was shown in Refs. \cite{Governale01, Storcz03,
vanderWal03}, the noise from the electromagnetic circuitry can be described
in terms of the impedance $Z(\omega )$ from a bath of $LC$ oscillators,
described by the Hamiltonian of interaction $H_{\mathrm{I}}=\frac{1}{2}%
\left( \sigma _{z}^{(a)}+\sigma _{z}^{(b)}\right) X$ in terms of the
collective bath coordinate $X=\sum\nolimits_{k}c_{k}\Phi _{k}$. Here $\Phi
_{k}$\ stands for the magnetic flux in the $k$-th oscillator, which is
coupled with the strength $c_{k}$\ to the qubits. It follows that the
relaxation tensor $\Gamma _{lmnk}$ is defined by the noise correlation
function $S(\omega )$%
\begin{eqnarray}
\Gamma _{lmnk} &=&\frac{\Lambda _{lmnk}}{\hbar ^{2}}S(\omega _{nk}),\text{ \
}S(\omega )=\int\limits_{0}^{\infty }dte^{-i\omega t}\left\langle
X(t)X(0)\right\rangle ,  \notag \\
\Lambda _{lmnk} &=&\left( \tau _{z}^{(1)}+\tau _{z}^{(2)}\right) _{lm}\left(
\tau _{z}^{(1)}+\tau _{z}^{(2)}\right) _{nk}.
\end{eqnarray}%
The correlator $S(\omega )$ was calculated in Refs.~\cite{Governale01,
Storcz03} within the spin-boson model and it was shown that the relevant
real part of the relaxation tensor
\begin{equation}
\text{Re}\Gamma _{lmnk}=\frac{1}{8\hbar }\Lambda _{lmnk}J(\omega _{nk})\left[
\coth \frac{\hbar \omega _{nk}}{2k_{B}T}-1\right]   \label{ReG}
\end{equation}%
is defined by the environmental Ohmic spectral density $J(\omega )=\alpha
\hbar \omega $ and is cut off at some large value $\omega _{\mathrm{c}}$,
where $\alpha $ is a parameter that describes the strength of the
dissipative effects.

From the above equations the expression for the relaxation rates from level $%
\left\vert n\right\rangle $ to level $\left\vert m\right\rangle $ follows%
\begin{equation}
W_{mn}=\frac{1}{4\hbar }\Lambda _{nmmn}J(\omega _{mn})\left[ \coth \frac{%
\hbar \omega _{mn}}{2T}-1\right] .  \label{Wmn}
\end{equation}%
In Ref.~\cite{Temchenko11} these relaxation rates were calculated as
functions of the partial flux biases $f_{a}$\ and $f_{b}$. It was
demonstrated that the fastest transitions are those between the energy
levels corresponding to changing the state of the first qubit and leaving
the same state of the second qubit. Such a difference in the relaxation
rates creates a sort of \textit{artificial selection rules} for the
transitions, similar to the selection rules studied e.g. in Refs.~\cite%
{Liu05, deGroot10, Niemczyk11}. To describe the hierarchy of the relaxation
rates, consider them in the simplified case, ignoring the interaction
between the qubits; then the single-qubit relaxation rates follow from Eqs.~(%
\ref{Wmn}) and~(\ref{gmn}) \cite{Blum, Chirolli08}%
\begin{eqnarray}
T_{1}^{-1} &=&W_{01}+W_{10}=\frac{\alpha \Delta ^{2}}{2\hbar \Delta E}\coth
\frac{\Delta E}{2T},  \label{T1} \\
T_{2}^{-1} &=&\text{Re}\gamma _{01}=\frac{1}{2}T_{1}^{-1}+\frac{\alpha T}{%
\hbar }\frac{\varepsilon ^{(0)2}}{\Delta E^{2}}.
\end{eqnarray}%
In particular, in the vicinity of the point $f_{b}=f_{b}^{\ast }$ in Fig.~%
\ref{Fig:levels_2qbs}(a), where $\Delta E^{(a)}=\Delta E^{(b)}$, we obtain $%
T_{1}^{(a)}/T_{1}^{(b)}\simeq \left( \Delta _{b}/\Delta _{a}\right) ^{2}$.
If $\Delta _{a}\gg \Delta _{b}$ is chosen, consequently the first qubit
relaxes much faster.

After the parametrization of the density matrix, $\rho _{ij}=x_{ij}+iy_{ij}$%
, the system's dynamics is described by the equations \cite{Temchenko11}
\begin{subequations}
\label{Eqs}
\begin{gather}
\dot{x}_{ii}=-\frac{1}{\hbar }\left[ V^{\prime },y\right] _{ii}+\sum_{r\neq
i}W_{ir}x_{rr}-W_{ii}x_{ii},\text{ }i=1,2,3; \\
\dot{x}_{ij}=\omega _{ij}y_{ij}-\frac{1}{\hbar }\left[ V^{\prime },y\right]
_{ij}-\gamma _{ij}x_{ij},\text{ }i>j; \\
\dot{y}_{ij}=-\omega _{ij}x_{ij}+\frac{1}{\hbar }\left[ V^{\prime },y\right]
_{ij}-\gamma _{ij}y_{ij},\text{ }i>j;
\end{gather}%
$y_{ii}=0$, $x_{00}=1-(x_{11}+x_{22}+x_{33})$; $x_{ji}=x_{ij}$, $%
y_{ji}=-y_{ij}$.

When discussing Fig.~\ref{Fig:levels_2qbs} we pointed out that in the system
of two coupled qubits there are two ways to realize lasing, making use of
the three or four levels to create the population inversion between the
operating levels. In Ref.~\cite{Temchenko11} the lasing in the two-qubit
system was demonstrated by solving numerically the Bloch-type equations (\ref%
{Eqs}). Besides demonstrating the population inversion between the operating
levels, an additional signal with the frequency matching the distance
between the operating levels was applied, to stimulate the transition from
the upper operating level to the lower one. So, the driving was considered
to be, first, the monochromatic signal $f(t)=f_{\mathrm{ac}}\sin \omega t$
to pump the system to the upper level and to demonstrate the population
inversion. Then another signal stimulating transitions between the operating
laser levels is applied $f(t)=f_{\mathrm{ac}}\sin \omega t+f_{\mathrm{L}%
}\sin \omega _{\mathrm{L}}t$ with $\hbar \omega _{\mathrm{L}}=E_{2}-E_{1}$.
Solving the system of equations (\ref{Eqs}), one obtains the population of $%
i $-th level of our two-qubit system, $P_{i}=x_{ii}$. The results of the
calculations are presented in Fig.~\ref{18}, where the temporal dynamics of
the level populations is given for two situations.

\begin{figure}[t]
\includegraphics[width=7cm]{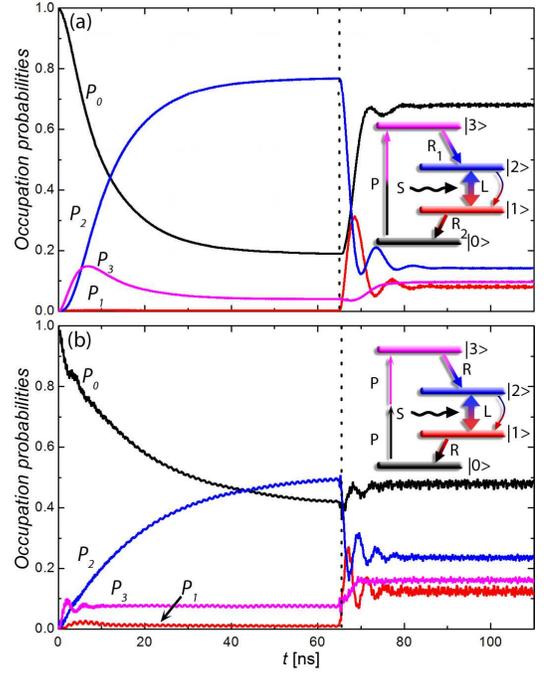}
\caption{(Color online) \textbf{Two-qubit lasing and stimulated transition}.
The time-dependent occupation probabilities are plotted for one- and
two-photon driving. The driving and fast relaxation create the inverse
population between the levels $\lvert {2}\rangle $ and $\lvert {1}\rangle $;
then the stimulating signal $f_{\mathrm{L}}\cos \protect\omega _{\mathrm{L}%
}t $ is turned on. \protect\cite{Temchenko11}}
\label{18}
\end{figure}


As shown in the inset schemes in Fig.~\ref{18}, the fastest (dominating)
relaxation transitions are $\lvert {3}\rangle \rightarrow \lvert {2}\rangle $
and $\lvert {1}\rangle \rightarrow \lvert {0}\rangle $. The system is
excited by either one- or two-photon transitions, with $\hbar \omega
=E_{3}-E_{0}$ in Fig.~\ref{18}(a) or with $2\hbar \omega =E_{3}-E_{0}$ in
Fig.~\ref{18}(b). This creates the population inversion between the levels $%
\lvert {2}\rangle \ $and $\lvert {1}\rangle $.\ Note that analogous
competition of the driving and relaxation can lead to the population
inversion in other multi-level systems \cite{Goorden05, Du10b}. Fast
relaxation, $\lvert {1}\rangle \rightarrow \lvert {0}\rangle $,\ helps
creating the population inversion between the laser levels $\lvert {2}%
\rangle \ $and $\lvert {1}\rangle $, which is the advantage of the
four-level scheme \cite{Svelto}. Then the transition $\lvert {2}\rangle
\rightarrow \lvert {1}\rangle $ is stimulated by another signal with a
frequency matching the laser operating levels ($\hbar \omega _{\mathrm{L}%
}=E_{2}-E_{1}$). Figure~\ref{18} was calculated for the following realistic
parameters \cite{Ilichev10}: $\Delta _{a}/h=15.8$ GHz, $\Delta _{b}/h=3.5$
GHz, $I_{\mathrm{p}}^{(a)}\Phi _{0}/h=375$ GHz, $I_{\mathrm{p}}^{(b)}\Phi
_{0}/h=700$ GHz, $J/h=3.8$ GHz, $k_{B}T/h=1$ GHz; and also $\omega _{\mathrm{%
L}}/2\pi =9$ GHz, $f_{\mathrm{L}}=f_{\mathrm{ac}}=5\times 10^{-3}$ with the
driving frequency $\omega /2\pi =47.4$ GHz for (a) and $\omega /2\pi =23.7$
GHz for (b).

For the realization of such lasing schemes, the system of two qubits should
be put in a quantum resonator, e.g. by coupling to a transmission line
resonator, as in Ref.~\onlinecite{Astafiev07}. Then the stimulated
transition between the operating states, demonstrated in Fig.~\ref{18}, will
result in transmitting the energy from the qubits to the resonator as
photons.

\section{Conclusions}

Here we presented the experimental and theoretical results of the study of
driven single and coupled superconducting qubits. The multiphoton
transitions in both charge and flux qubits were studied in details. Those
processes are important for both demonstrating the fundamental quantum
phenomena in mesoscopic systems and for developing controlling mechanisms
for perspective devices.

The system of qubits, coupled to the controlling electronics and measuring
resonator, can be described within the semiclassical approach. After
presenting this formalism in application to probing the qubit systems, we
have shown some specific experimental results, which were accompanied by the
calculated counterparts. The agreement between them shows contemporary
possibility to demonstrate and describe quantum phenomena in mesoscopic
systems.

\begin{acknowledgments}
The results presented here were obtained together with many our colleagues
who were our co-authors in the respective publications. We are grateful to
all of them for their contributions. We thank S. Ashhab for useful comments.
SNS acknowledges the hospitality of IPHT during his visit. This work was
partly supported by NAS of Ukraine (Project No. 04/10-N), DKNII (Project No.
M/411-2011), BMBF (UKR 10/001), EU project (IQIT).
\end{acknowledgments}

\end{subequations}

\end{document}